\documentclass[a4paper,fleqn,usenatbib]{mnras}

\usepackage{newtxtext,newtxmath}

\usepackage[T1]{fontenc}
\usepackage{ae,aecompl}

\usepackage{ifpdf}

\usepackage{graphicx}
\usepackage{amsmath}
\usepackage{amssymb}

\usepackage[dvipsnames]{xcolor}
\usepackage[abs]{overpic}

\def\msun{{\rm M}_\odot}
\def\lsim{\mathrel{\rlap{\lower 3pt \hbox{$\sim$}} \raise 2.0pt \hbox{$<$}}}
\def\gsim{\mathrel{\rlap{\lower 3pt \hbox{$\sim$}} \raise 2.0pt \hbox{$>$}}}

\title[Bar formation triggers]{External versus internal triggers of bar
  formation in cosmological zoom-in simulations}

\author[T. Zana et al.]{Tommaso Zana,$^{1}$\thanks{E-mail: tzana@studenti.uninsubria.it} Massimo Dotti,$^{2,3}$ Pedro~R. Capelo,$^{4}$ Silvia Bonoli,$^{5}$ \newauthor Francesco Haardt,$^{1,3}$ Lucio Mayer$^{4}$ and Daniele Spinoso$^{5}$\\
$^{1}$DiSAT, Universit\`a degli Studi dell'Insubria, Via Valleggio 11, I-22100 Como, Italy\\
$^{2}$Dipartimento di Fisica G. Occhialini, Universit\`a di Milano-Bicocca,
Piazza della Scienza 3, I-20126 Milano, Italy\\
$^{3}$INFN, Sezione di Milano-Bicocca, Piazza della Scienza 3, I-20126 Milano,
Italy\\
$^{4}$Center for Theoretical Astrophysics and Cosmology, Institute for 
Computational Science, University of Zurich,\\
Winterthurerstrasse 190, CH-8057 Z$\ddot{u}$rich, Switzerland\\
$^{5}$Centro de Estudios de F{\'i}sica del Cosmos de Arag{\'o}n, plaza San Juan, 1 planta-2 E-44001 Teruel, Spain\\
}

\date{Accepted 2017 September 24. Received 2017 September 5; in original form 2017 April 26}

\pubyear{2017}

\begin{document}
\label{firstpage}
\pagerange{\pageref{firstpage}--\pageref{lastpage}}
\maketitle

\begin{abstract}
The emergence of a large-scale stellar bar is one of the most striking features in disc galaxies.
By means of state-of-the-art cosmological zoom-in simulations, we study the formation and evolution of bars in Milky Way-like galaxies in a fully cosmological context, including the physics of gas dissipation, star formation, and supernova feedback. 
Our goal is to characterise the actual trigger of the non-axisymmetric perturbation that leads to the strong bar observable in the simulations at $z=0$, discriminating between an internal/secular versus an external/tidal origin. To this aim, we run a suite of cosmological zoom-in simulations altering the original history of galaxy-satellite interactions at a time when the main galaxy, though already bar-unstable, does not feature any non-axisymmetric structure yet. We find that the main effect of a late minor merger and of a close fly-by is to delay the time of bar formation and those two dynamical events {\it are not} directly responsible for the development of the bar and do not alter significantly its global properties (e.g. its final extension). We conclude that, once the disc has grown to a mass large enough to sustain global non-axisymmetric modes, then bar formation is inevitable. 
\end{abstract}

\begin{keywords}
methods: numerical -- galaxies: evolution -- galaxies: kinematics and
dynamics -- galaxies: structure
\end{keywords}



\section{Introduction}
 
The frequent occurrence of stellar bars in disc galaxies\footnote{$\gsim 30$ per cent of massive galaxies \citep{laurikainen, nair, lee, gavazzi, consolandiB}.} has promoted a wealth of observational and theoretical studies focussed on the effects that such non-axisymmetric structures can have on the galactic constituents, including pre-existing \citep[e.g.][]{lutticke, bureau, kormendy} and newly formed stars \citep{ho, martinet, hunt, laurikainen, jogee}, as well as diffuse gas \citep[e.g.][]{sanders, roberts, athanassoula, cheung, fanali, hakobyan} and dust \citep[e.g.][]{consolandi}.
Moreover, bars are thought to be at the origin of the frequently observed \citep{lutticke} boxy/peanut-shaped bulges (\citealt{combes}; B/P from now on) and of the X-shaped bulges, when the bar growth continues undisturbed and the peanut
shape becomes more pronounced \citep{athanassoulaC}.

Several processes have been proposed as drivers of bar formation: a bar can either grow from small non-axisymmetric perturbations within the host \citep[e.g.][]{hohl, ostriker, sellwood} or can be triggered  by interactions with external perturbers, such as galaxies in mergers and close fly-by encounters \citep[e.g.][]{byrd, mayer, curir, gauthier, romano, martinez} and galaxy clusters \citep{lokas}. 
Recent observational investigations reached contrasting results regarding the role of environment in the bar formation process. Some studies indeed found an excess of barred systems in dense galactic environments \citep[e.g.][]{skibba}, whereas other works support the idea that dynamical perturbations have the primary effect of suppressing the formation of a bar \citep{li, lin}.
The relative importance of the different triggers of bar growth is still unclear, as it is extremely difficult to identify the very early stages of bar formation (when a clear structure is not present yet).
From a theoretical point of view, cosmological simulations can help in understanding which processes led to the growth of bars. Indeed, in numerical experiments the whole 3D distribution of matter is completely characterized at any needed time and it is, in principle, possible to study the effect of every single interaction on every galaxy undergoing bar instability \citep[e.g.][]{moetazedian}. Unfortunately, the number of studies that focus on the study of bars in a fully cosmological perspective, including the growth of the host galaxy through a sequence of mergers as well as accretion of diffuse gas, is still limited, as they require high spatial resolution ($\lsim 100$ pc, in order to properly resolve the stellar dynamics within the bar region) while evolving a large cosmological box \citep{romano, kraljic, scannapieco, goz, okamoto, spinoso, algorry, sokowlowska}.

Our study focusses on the analysis of several replicas of the ErisBH simulation \citep{bonoli}, a cosmological zoom-in run that led to the formation of a realistic Milky Way-like galaxy hosting a bar in its central few kpc. The properties of the bar and its effect on to the host galaxy and its gas and stellar component have been detailed in \cite{spinoso}. Interestingly, the gravitational potential of the main galaxy becomes prone to bar instability in its central region before the last significant merger (with a mass ratio $\gsim 0.05$), but develops an observable bar only at later stages \citep{spinoso}. The susceptibility to bar formation was convincingly inferred via the combination of two criteria based on the Toomre $Q$ and swing amplification $X$ parameters \citep[for the $m=2$ mode;][]{toomre, toomreB}. In collisionless systems, which cannot dissipate energy, these two criteria combined are known to express a necessary and sufficient condition for bar formation \citep[e.g.][]{binney}. The aim of our investigation is to identify the actual trigger of the bar formation, in particular whether or not the disc's own self-gravity and structure are responsible for it, or if is instead crucial the role of external perturbations (e.g. infalling satellites).

In order to do so, we analysed the growth history of ErisBH's most massive galaxy by identifying the most relevant mergers and fly-bys. We then ran a suite of ad-hoc simulations, either removing these interacting galaxies or changing their parameters (orbital or structural), checked whether a bar would form and, if so, we studied the details of its growth. The rationale behind ``engineering" the simulations by changing a targeted aspect of the initial conditions is similar to the approach followed by \cite{pontzen}, who applied small changes in the initial conditions
of zoom-in simulations in order to produce different accretion histories leading to the same final mass and environmental properties.

This paper is organized as follows: we summarise the properties of the original ErisBH in Section \ref{subsec:ICeris}. In Section \ref{subsec:ICsiblings}, we describe the parameters of the new runs performed, highlighting the differences with respect to the parent simulation. In Section \ref{sec:result}, we discuss the outcomes of our investigations, and in Section \ref{sec:conclusion} we present a discussion of the results and draw our conclusions.

\section{Initial conditions}
\label{sec:IC}

\begin{figure*}
\begin{overpic}[width=0.91\textwidth]{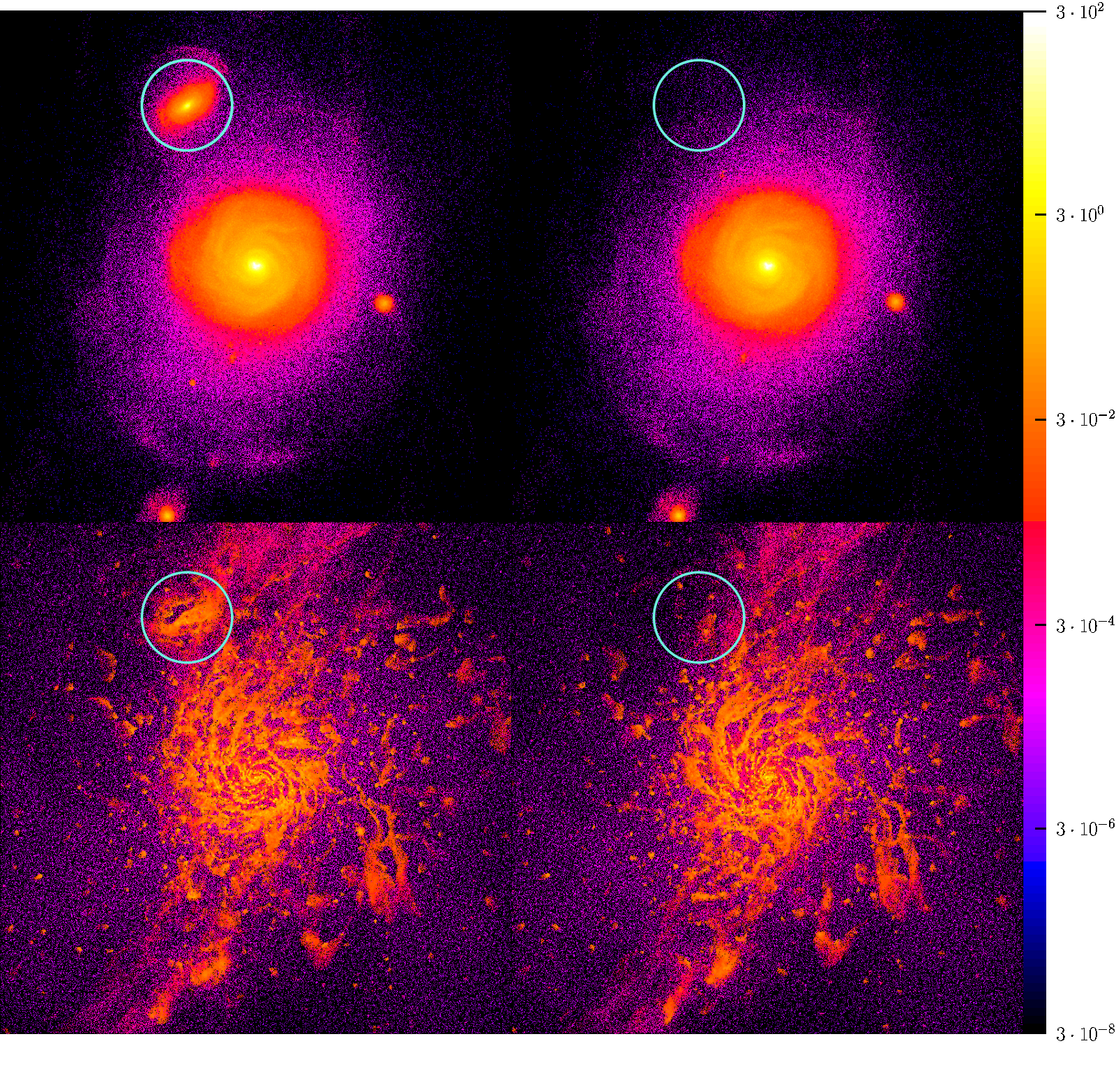}
\put(440,232){\colorbox{white}{\parbox{0.9cm}
	{
	\rotatebox{90}{$\left[\msun \cdot \rm{pc}^{-3}\right]$}
    }}}
\end{overpic}
\vspace{-0.5cm}
\caption{Upper panels: stellar density map of the central 50 kpc region for the WM (left panel) and NM (right panel) runs at $z \approx 1.7$. The colour scale is logarithmic spaced with minimum value 3 $\times \, 10^{-8} \, \msun \rm{pc}^{-3}$ and maximum value $300 \, \msun \rm{pc}^{-3}$. Lower panels: same as the upper panels, for the gas distribution. The original location of the companion that merges at $z \approx 1.2$ in the WM run is highlighted in each panel with a circle of radius $\sim 5$ kpc.} 
\label{fig:merger_removal}
\end{figure*}

\subsection{The parent run: ErisBH}
\label{subsec:ICeris}

Eris \citep{guedes} and ErisBH \citep{bonoli} are two smoothed-particle hydrodynamics (SPH) zoom-in simulations of the same Milky Way-sized halo selected within a low-resolution, dark matter-only simulation of a comoving cube of $(90$ Mpc$)^3$. Both runs assumed a flat Universe with $\Omega_{\rm M} =$ 0.24, $\Omega_{\rm b}=$ 0.042, $h_{\rm 0} =$ 73 km s$^{-1}$ Mpc$^{-1}$, and $\sigma_{8} =$ 0.76, as derived from the \textit{Wilkinson Microwave Anisotropy Probe} three-year data \citep{spergel}. A comoving region of $(1$ Mpc$)^3$ has been re-sampled at higher resolution with 13 million dark matter particles of mass $m_{\rm DM}= 9.8 \,\times 10^4 \msun$ and populated with additional 13 million SPH particles. The whole system has been evolved from $z=90$ to $z=0$ using the tree-SPH code {\scshape gasoline} \citep{wadsley}. The gravitational softening for each particle, fixed to 120 physical pc in the $0<z<9$ range, evolves $\propto (1+z)^{-1}$ at earlier times.

Both Eris and ErisBH included Compton and atomic cooling of primordial gas, heating from a UV background \citep{haardt}, and metallicity-dependent radiative cooling at low ($<10^4 \rm K$) gas temperatures (see \citealt{guedes} for more details). Gas particles belonging to a converging flow that cross the star-formation temperature ($T_{\rm SF}=3 \times 10^4$ K) and density ($n_{\rm SF}=5$ atoms cm$^{-3}$) thresholds can form stars of initial mass $m_{*}= 6 \,\times 10^3 \msun$, each representing a stellar population described by a \cite{kroupa} initial stellar mass function. Each supernova (SN) exploding releases $8 \times 10^{50}$ erg to the neighbouring gas particles following the blastwave prescription in \cite{stinson}.

ErisBH differs from Eris by including prescriptions for the formation and growth of massive black holes (MBHs) and the feedback that these can exert on to the galactic medium during accretion events \citep{bellovary, bonoli}. In particular, a MBH can be seeded in a halo only when: $(i)$ the halo does not host any other MBH, $(ii)$ the halo is properly resolved (by at least 100k particles), and $(iii)$ it hosts a
dense gas structure with more than 10 particles exceeding a density of 100 atoms cm$^{-3}$. Every MBH can accrete mass from the surrounding medium following the Bondi--Hoyle--Lyttleton prescription \citep{hoyle, bondi, bondiB}:

\begin{equation}
\dot{M}_{\rm Bondi} = \frac{4 \pi G^2 M^2_{\rm BH} \rho}{(c^2_{\rm s}+v^2)^{3/2}}, 
\label{eq:bondi}
\end{equation}
where $G$ is the gravitational constant, $M_{\rm BH}$ is the mass of the MBH, and $\rho$, $c_{\rm s}$, and $v$ are the gas density, sound speed, and relative velocity with respect to the MBH, evaluated over the closest 32 SPH particles, respectively. The maximum accretion rate is capped at the Eddington limit,

\begin{equation}
\dot{M}_{\rm Edd} = \frac{4 \pi G M_{\rm BH} m_{\rm p}}{\eta \sigma_{\rm T} c},
\label{eq:edd}
\end{equation}
that we computed assuming a radiative efficiency of $\eta = 0.1$. Here, $m_{\rm p}$ is the proton mass, $\sigma_{\rm T}$ the Thomson cross section, and $c$ the speed of light in vacuum.
The same efficiency is assumed when computing the radiated luminosity, a fraction $\epsilon_f=0.05$ of which is 
coupled to the 32 gas neighbours as a thermal energy injection (see \citealt{bellovary} for a full detailed description of the implementation of the MBH physics).

We stress that the MBH feedback has an impact on to the ErisBH galaxy at early times ($z \gsim 2$), in preventing the formation of a central baryonic overdensity, so that at lower redshifts the galactic potential can sustain a central bar
(contrary to the Eris case, as discussed in \citealt{spinoso}; see also \citealt{rodionov}). However, the effect of MBH growth at lower redshifts is essentially negligible, as the central object grows at very low rates, for a total accreted mass of only $14$ per cent of its final value since $z\approx 1.2$. For these reasons, in our investigation we turned off accretion and feedback from the central MBH.

As fully detailed in \cite{spinoso}, a strong bar is clearly observable in the centre of the ErisBH disc galaxy at $z=0$. The bar starts growing by $z \lsim 1$ and it is unclear whether the trigger of its growth could be associated to a minor merger occurring at $z\approx 1.2$, that indeed imprints a clear non-axisymmetric perturbation in the (already bar-unstable) central region of
the galaxy. The merging satellite has a stellar mass of $M_{*} \approx 1.2 \times 10^{9} \msun $ and a dark component mass of $M_{\rm DM} \approx 1.5 \times 10^{10} \msun $, to be compared to the dominant galaxy which has $M_{*} \approx 2.0 \times 10^{10} \msun $ and $M_{\rm DM} \approx 1.1 \times 10^{11} \msun$ within a 20 kpc radius from the centre.\footnote{The reported figures are estimated at $ z \approx 1.7$, hence before the merger.} Other minor perturbations are present at lower redshifts, the most significant being multiple fly-bys of a relatively massive satellite at $z\approx 0.65$, $0.35$, and $0.15$. The stellar mass of such satellite is $M_{*} \approx 2.1 \times 10^{8} \msun$, whereas that of its dark matter component is $M_{\rm DM} \approx 1.4 \times 10^{9} \msun$. These values are computed shortly before the first closest-approach point, at about $48.1$ kpc from the main galaxy.

\subsection{The siblings}
\label{subsec:ICsiblings}

Taking full advantage of the ErisBH run, in order to isolate the influence of various dynamical events over bar formation and evolution processes, we produced five slightly different initial-condition files, starting from a snapshot at $z \approx 1.8$ in the evolutionary history of the original simulation.

All five runs share a number of differences with respect to ErisBH. In particular, (i) we evolved a smaller number of particles (we cut the system at $4500$ comoving kpc from the centre of mass of the dominant galaxy), (ii) the BH accretion was switched off to further simplify the setup, and (iii) we could not retrieve whether the gas particles were recently subject to SN feedback.

Since these peculiarities may prevent an accurate comparison between ErisBH and its modified replicas,
one of the five runs we evolved is a twin simulation of ErisBH ({\bf run WM} hereon), though with the differences noted above. 
This run allows us to check how the
aforementioned differences bias the evolution of the system compared to ErisBH. In addition, we stress that this paper will be focussed on the comparison between the siblings described below and run WM rather than ErisBH itself, as in this manner we are sure to cancel out the variations introduced at restart.

The other four runs were re-initialized modifying the initial conditions at $z\approx 1.8$, when the galaxy is already prone to bar instability but no central asymmetries are visible \citep[see Figure~3 in][]{spinoso}, as follows.

\noindent In {\bf run NM}, we removed only the satellite which merges at $z \approx 1.2$ in run WM. In order to prevent numerical effects and to maintain the self-consistency of the simulation, we removed the object when it was $40.1$ kpc away from the galaxy at $z \approx 1.8$. Figure~\ref{fig:merger_removal} shows a comparison between the stellar and gas density maps for the runs WM and NM in the vicinity of the primary galaxy, at $z\approx 1.7$.

\begin{figure*}
\begin{overpic}[width=\textwidth]{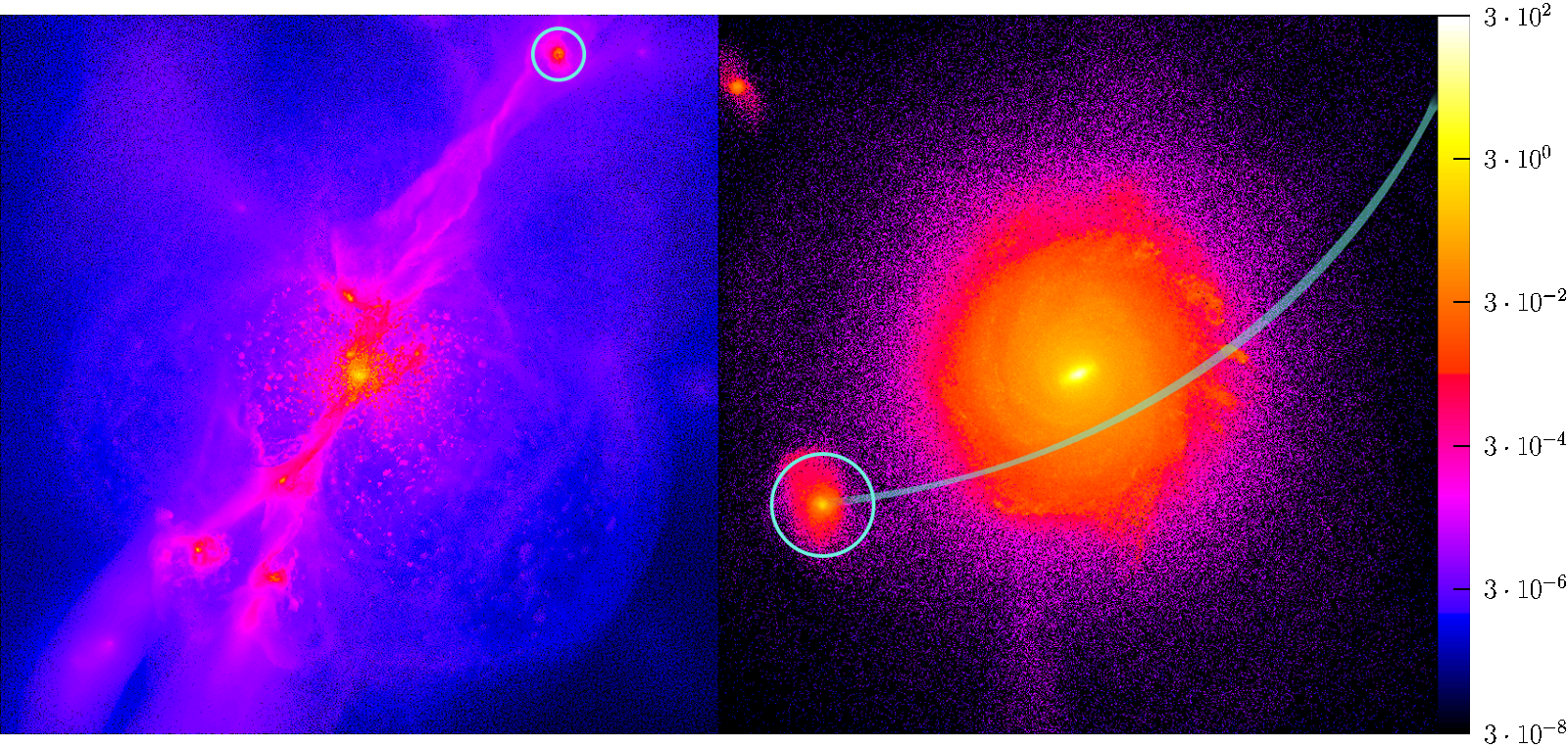}
\put(482,118){\colorbox{white}{\parbox{0.3cm}
	{
	\rotatebox{90}{\tiny $\left[\msun \cdot \rm{pc}^{-3}\right]$}
    }}}
\end{overpic} 
\caption{Left panel: gas density map for the WM run at $z \approx 1.7$ in a box of physical side 554 kpc. Right panel: stellar density map of the central 55 kpc from the same run, immediately after the first pericentric passage of the fly-by ($z \approx 0.65$). The colour scale is the same applied in Figure~\ref{fig:merger_removal}. The position of the fly-by location and its trajectory during the closest approach are marked in cyan.}
\label{fig:flyby}
\end{figure*}

\noindent The second largest perturbation to the primary galaxy consists of a fly-by, i.e. an object which in the WM run performs three close passages near the central galaxy at $z \lsim 0.7$. The position of this satellite at $z \approx 1.7$ is highlighted in the left panel (density map of the gas component) of Figure~\ref{fig:flyby}, whereas a zoom-in and its trajectory during the first pericentre are shown in the right panel of the same figure (stellar density map). {\bf Run NF} is a copy of run NM, except that we also removed the fly-by, forcing a dynamical evolution with no significant events after $z \approx 1.2$.

\noindent In {\bf run PF}, we allowed the formation of the fly-by satellite, but we provided it with a velocity radial component in order to push the object away from the central galaxy. This method makes the fly-by satellite reach a distance from the main galaxy as large as $\approx 500$ kpc at $z \approx 0.5$, thus reducing its influence over the main system.

\noindent Finally, in {\bf run HF}, the orbital parameters of the fly-by were kept as in run NM, 
but we doubled the mass of its dark matter component. 
The main characteristics of our five runs are summarised in Table~\ref{tab:example_table}.

The described initial conditions were evolved for a total of approximately 1.0 million CPU hours via the tree-SPH code {\scshape ChaNGa} \citep{menon}, a program for cosmological simulations built on the tree-SPH code {\scshape gasoline} but implemented with the parallel programming system Charm++. Thanks to Charm++'s dynamic load balancing algorithms, {\scshape ChaNGa} allows to obtain high performances over huge dynamical {\it and} mass ranges \citep[for further technical details, see][]{jetley}.

We investigated a temporal window from $z \approx 1.8$ down to $z=0$, using the same cosmological parameters adopted for the Eris suite. The selected volumes host a total number of particles of about $3 \times 10^{7}$, subdivided in $1.2 \times 10^{7}$ SPH, $1.3 \times 10^{7}$ dark matter, and $4 \times 10^{6}$ star particles.

In order to preserve continuity with the parent run, both Compton and atomic hydrogen cooling are included in the adopted code.
Moreover, the energetic state of the gas takes into account the same UV background considered in ErisBH \citep{haardt}. The process of star formation and SN feedback follows again the method described in \citeauthor{stinson} (\citeyear{stinson}; efficiencies and other parameters remained unchanged, except for the IMF model, which is in our case that of \citealt{kroupa93}). However, no prescription for BH accretion and feedback was implemented. The gravitational softening in our suite was kept fixed to $120$ comoving pc, implying that 
the {\it physical} softening decreases for increasing redshift.

\begin{table}
\centering
\caption{Summary of the runs.}
\label{tab:example_table}
\begin{tabular}{ccc}
	\hline	
	Run & Extended Name & Details \\
	\hline \hline
	WM & With Merger & control run\\
	\hline
	NM & No Merger & no minor merger\\
	\hline		
	NF & No Fly-by & no minor merger \& no fly-by\\
	\hline		
	PF & Pushed Fly-by & no minor merger \& fly-by pushed\\
	\hline
	HF & Heavy Fly-by & no minor merger \& heavier fly-by\\
	\hline
\end{tabular}
\end{table}

\section{ANALYSIS AND RESULTS}
\label{sec:result}

\begin{figure}
\includegraphics[width=\columnwidth]{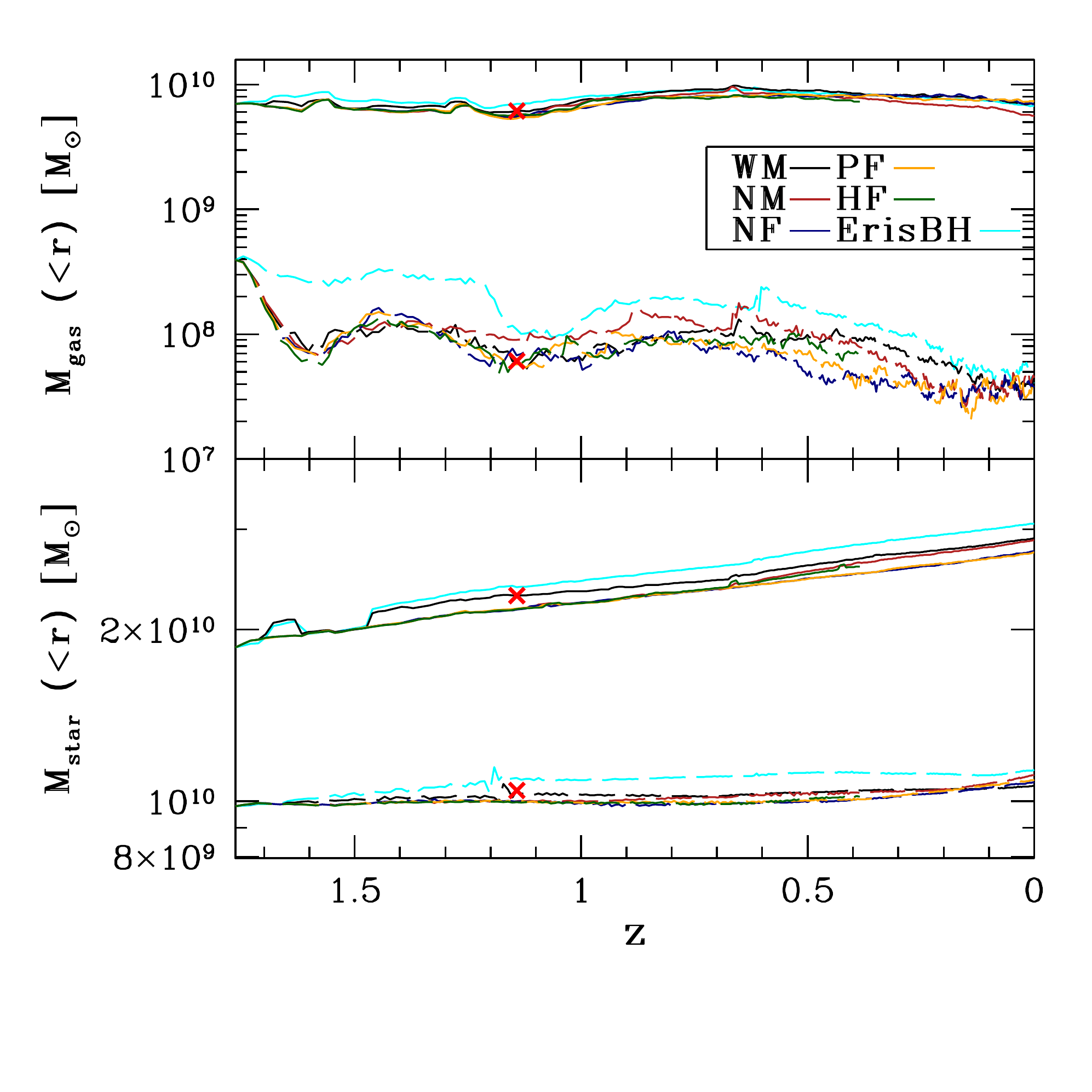}
\vspace{-1cm}
\caption{Gas (upper panel) and stellar (lower panel) mass enclosed within spheres of 20 (solid lines) and 2 (dashed lines) physical kpc radii. The comparison with the ErisBH original run is provided by the cyan line. The effect of the last minor merger in the ErisBH twin run (run WM; black line) can be observed in the sudden stellar mass growth observable at $z\approx 1.5$, when the satellite crosses the 10 kpc boundary. The completion of the merger occurs at $z\approx 1.2$ (this last event is marked with the red cross).}
\label{fig:mass}
\end{figure}

The main goal of our analysis is to assess the actual trigger of the bar instability (external-tidal versus internal-secular mechanism) in our simulation suite. Hence, we start by checking whether significant differences amongst the five runs, other than the properties of the satellites, actually exist. Since in four out of five runs we removed the last minor merger experienced by the central galaxy of ErisBH at $z\approx 1.2$, the mass of the primary galaxy is not exactly the same in all runs. Figure~\ref{fig:mass} quantifies the redshift evolution of the primary galaxy mass in the different runs considered. 
Small mass variations are indeed present, still the maximum relative difference in stellar mass is $\lsim 7$ per cent at both small and large scales, with the largest difference observable in correspondence of the minor merger (the stellar mass of the removed satellite represents $\simeq 6$ per cent of the main galaxy stellar component at $z\approx 1.7$; this value drops to $\simeq 0.9$ per cent by $z=0$). The gas mass within 20 kpc shows larger variations (about $10$ per cent between the different runs).\footnote{The gas mass within 2 kpc shows even larger variations at late times. Differences at such small scales are not related to gas-replenishment episodes driven by mergers/fly-bys, nor to different properties of the large-scale gas streams fuelling the primary; rather, they stem from the difference in the formation time of the bar.}
We stress, however, that the gas mass never exceeds $40$ per cent ($5$ per cent) of the stellar mass within 20 (2) kpc. Star formation from gas infalling through large-scale cold streams (clearly observable in the left panel of Figure~\ref{fig:flyby}) is responsible for the galaxy mass growth during the last $\approx9.7$ Gyr, while mergers with smaller structures are a second-order perturbation to the mass content of the galaxy for $z\lsim 1.5$. The only notable difference with respect to ErisBH (cyan line in Figure~\ref{fig:mass}) is represented by the initially larger gas mass within 2 kpc. The effect can be explained by considering that our new simulations start without gas cooling shut-off by previously exploded SNae, so that the resulting higher star formation rate leads to a rapid consumption of the gas reservoirs.\footnote{We stress that, although we often show direct comparisons of our results with ErisBH, it is more meaningful to compare them to run WM, given the actual differences between these two simulations discussed in Section~\ref{subsec:ICsiblings}.}

\subsection{Fourier analysis}

\begin{figure*}
\includegraphics[width=17.55cm]{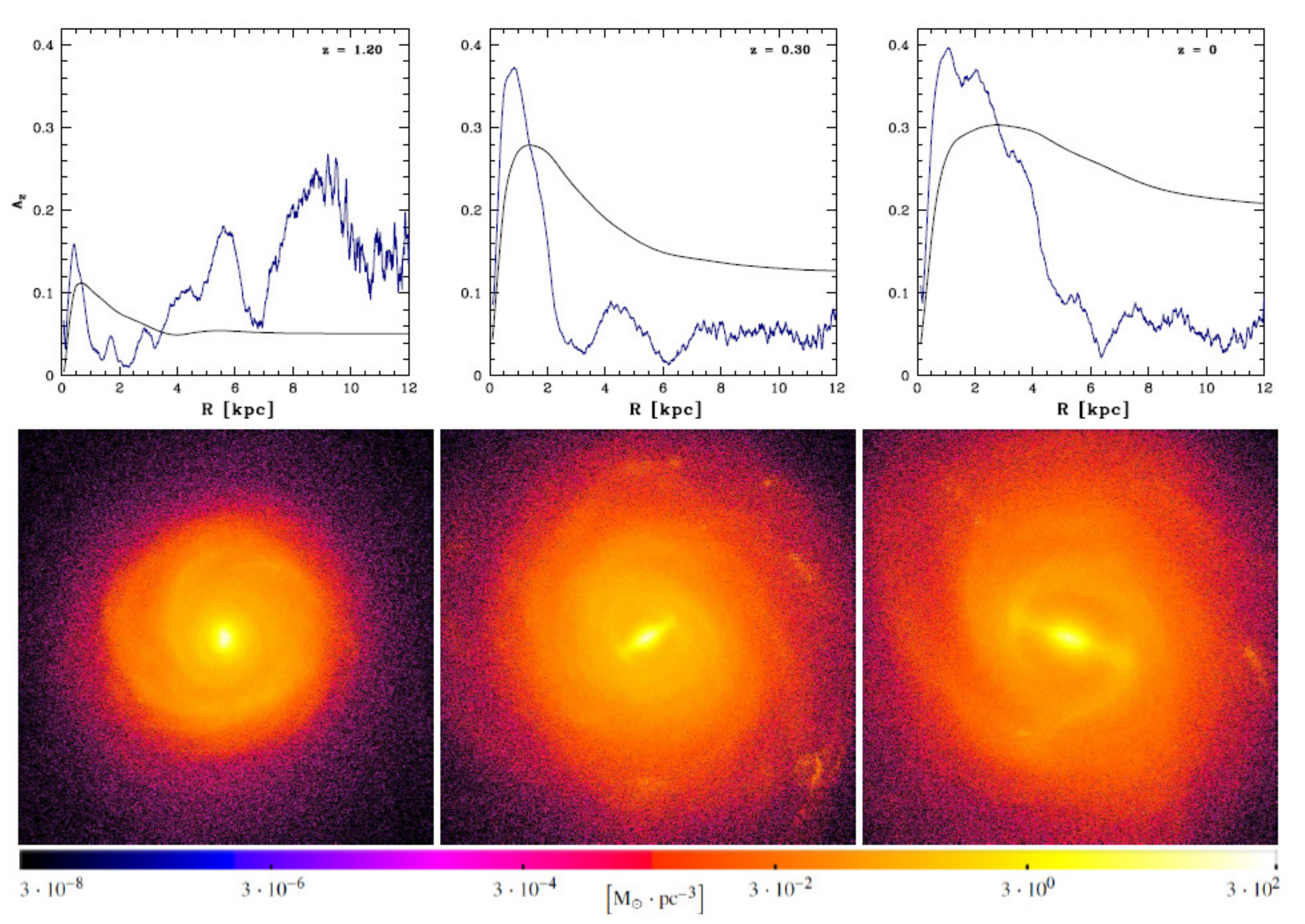}
\caption{The evolution of the $A_{2}(R)$ profile (blue line) and its cumulative counterpart $A_{2}(<R)$ (black line) of the primary galaxy in run NM at three different redshifts (upper panels) is reflected in the corresponding stellar density maps (lower panels) of the central 25 kpc, with the same colour coding as in Figures~\ref{fig:merger_removal} and \ref{fig:flyby}. The major peak of $A_{2}(R)$ at $R \approx 9$ kpc at $z=1.2$ is due to the lower density of the outskirts of the galactic disc.}
\label{fig:examples}
\end{figure*}

\begin{figure*}
\includegraphics[width=\textwidth]{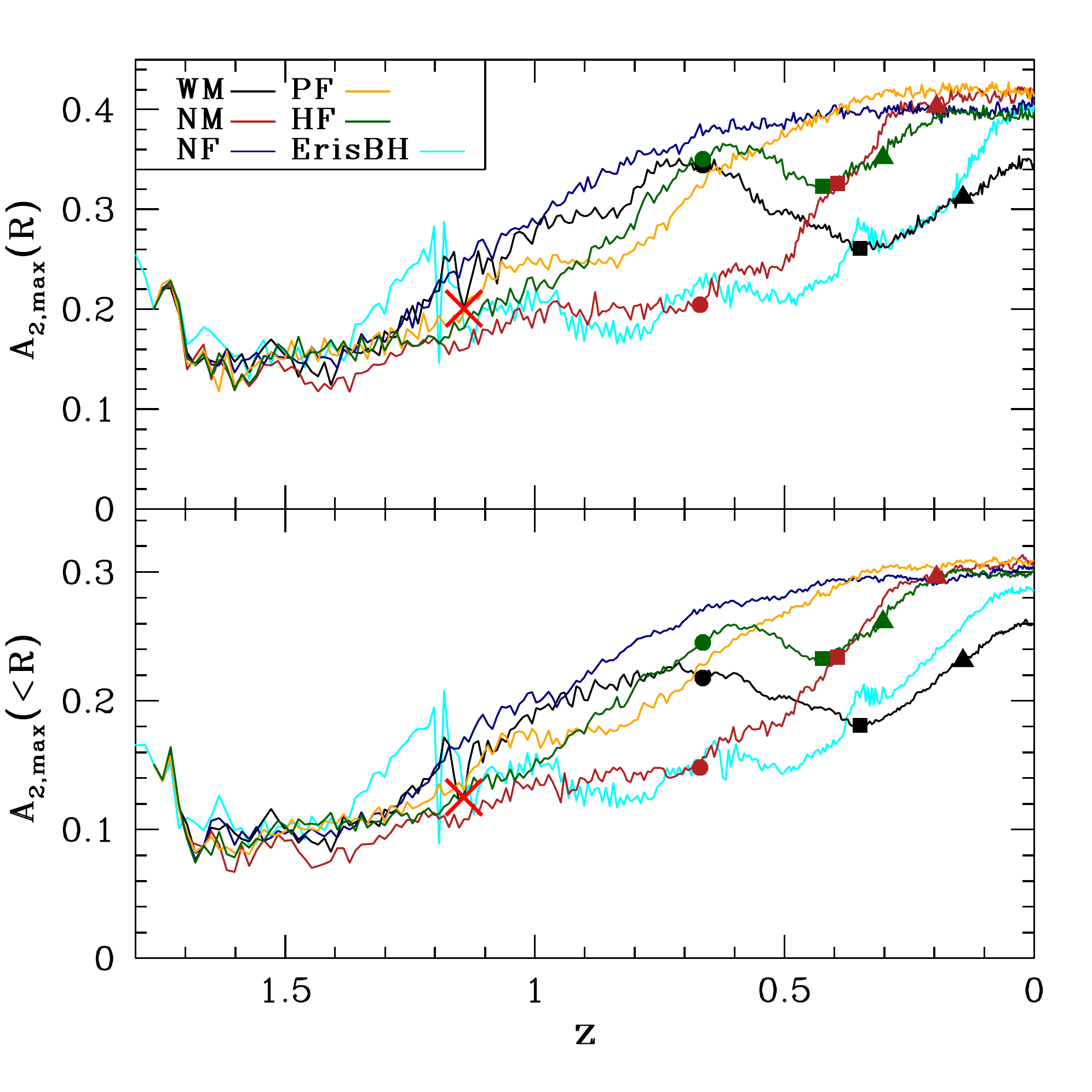}
\vspace{-0.5cm}
\caption{Redshift evolution of the bar intensity for our five runs, compared with the ErisBH original simulation. The quantities $A_{2, \rm{max}}(R)$ (upper panel) and $A_{2, \rm{max}}(<R)$ (lower panel) show almost the same trend, proving that the study of the bar growth is not biased by the central mass concentration. The time at which the minor merger occurs in run WM is highlighted with a red cross, whereas the times of the first, second, and third pericentre passages of the fly-by are marked as circles, squares, and triangles, respectively.}
\label{fig:A2}
\end{figure*}

\begin{figure*}
\includegraphics[width=\textwidth]{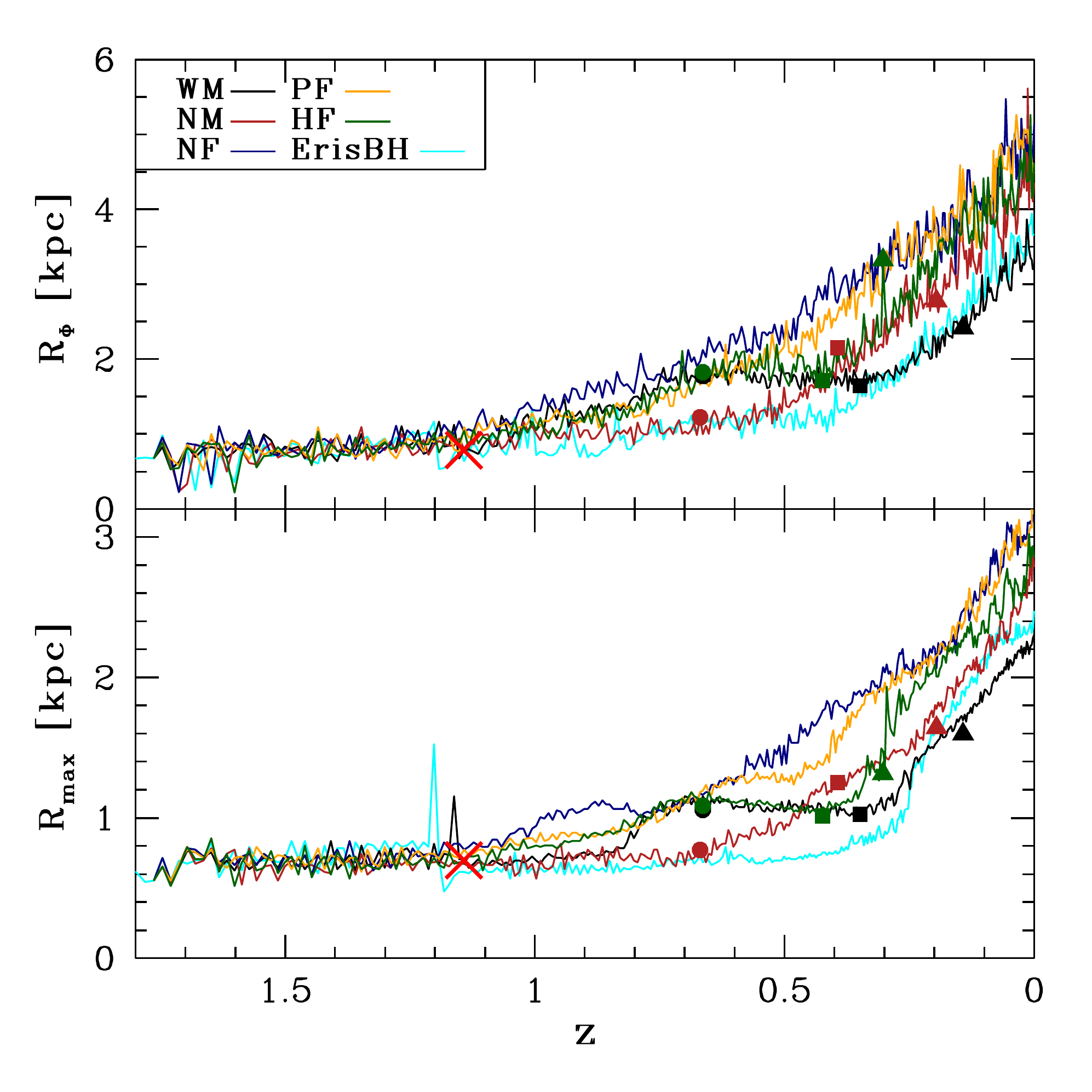}
\vspace{-0.5cm}
\caption{Bar extent as a function of redshift: the upper panel shows the evolution of $R_{\Phi}$, whereas $R_{\rm{max}}$ is shown in the lower panel. The colour code and the symbol legend are the same adopted in Figure~\ref{fig:A2}. The agreement between the trend of the two methods is excellent, though $R_{\Phi}$ is higher at all redshifts.}
\label{fig:R}
\end{figure*}

Once we checked that all the main galaxies share a similar mass evolution, we assess whether different sequences of interactions with the satellites would result in different morphologies of the primary galaxy. 

\subsubsection{Bar strength}
\label{ss:bar_stength}

We quantify the presence (or absence) and the intensity of a central bar by performing the Fourier decomposition of the stellar surface density field (on a face-on projection). We then evaluate the ratio between the second term of the expansion, which captures the strength of two-fold structures, and the zeroth (axisymmetric) term through the relation

\begin{equation}
A_2(<R) \equiv \frac{\left|\sum_{j}{m_{j}e^{2i\theta_{j}}}\right|}{\sum_{j}m_{j}} ,
\label{eq:A2}
\end{equation}
where $m_{j}$ is the mass of the $j$-th particle and $\theta_{j}$ the azimuthal angle of its projection over the disc plane. The summation is carried over all particles enclosed within a cylinder of radius $R$, coaxial with the galaxy, and of fixed height of $1$ kpc. $A_2(<R)$ increases as far as the stellar distribution is decisively non-axisymmetric, decreasing gradually farther out.

Even though, for our purposes, the bar strength appears to be well represented by the trend of the $A_{2}$ cumulative profile (see Figure~\ref{fig:examples}), it is worth noting that its value could depend on the central mass concentration of the galaxy. For this reason, we provide here also the result of the differential Fourier analysis, where we first divide the disc into annuli of equal width, and then we compute the value of the ratio $A_{2}(R)$ separately for each annulus at radius $R$ \citep[similar studies have been performed by e.g.][]{athanassoulaB, valenzuela}. In the Fourier decomposition we consider only stars within a $1$ kpc height.

Figure~\ref{fig:examples} shows for the NM run at three different redshifts the $A_{2}(R)$ profile (blue line) and its cumulative equivalent (black line) as a function of the cylindrical radius.\footnote{In this work we use $A_2(<R)$ to indicate the cumulative values, whereas $A_2(R)$ refers to the differential one.} Each plot is accompanied by the corresponding face-on stellar density maps.
The plot clearly shows the non-axisymmetric nature of the galaxy disc at three different epochs along the galaxy evolution. 
It is clear how a two-fold overdensity is evolving, gaining both amplitude and radial extent as the galaxy approaches $z=0$.
This evolution is evident in both lines even if it is also visible that the central density slightly biases the trend of $A_{2}(<R)$ at larger radii.
Nevertheless, it is worth noting that we are mainly interested in the innermost kpc of the galactic disc, where the bar actually forms, and here both the local maxima of $A_{2}(<R)$ and $A_{2}(R)$, $A_{2, \rm{max}}(<R)$ and $A_{2, \rm{max}}(R)$, respectively, provide a reliable estimate of the current strength of the growing non-axisymmetry.
The values of $A_{2}(R)$ have been computed over a much smaller number of particles (each annulus is just a fraction of the total disc volume), thus even a slight variation in symmetry is highlighted in the final result. As a consequence, the differential function fluctuates considerably more than the cumulative one.

The evolution of $A_{2, \rm{max}}(R)$ and $A_{2, \rm{max}}(<R)$ is shown in Figure~\ref{fig:A2} for all runs, as well as for ErisBH.

\subsubsection{Bar length}

A very important part in the study of galactic bars is the evaluation of their length. This is a very debated subject, since the body of the bar has no perfectly defined boundaries. Moreover, it is extremely difficult to match numerical results with observations, especially because many approaches used to evaluate such property cannot be adopted in both these two fields of study \citep[for further details, see e.g.][]{debattistaB, athanassoulaB}.

In the present work, we utilize a method to measure the bar length described in \cite{athanassoulaB}. A by-product of the Fourier decomposition is represented by the phase of the $A_{2}(R)$ component, defined as:

\begin{equation}
\Phi(R) \equiv \frac{1}{2} \arctan \left[ \frac{\sum_{j}{m_{j}\sin(2\theta_{j})}}{\sum_{j}{m_{j}\cos(2\theta_{j})}} \right] ,
\label{eq:phase}
\end{equation}
where the summation is performed along each annulus of radius $R$.
Since this phase is nearly constant in the range of the bar overdensity\footnote{The fluctuations in this case may depend on the mass resolution of the numerical simulation.} and is randomly oriented outside, the length of the bar can be defined as the radius $R_{\Phi}$ of the outmost annulus with phase constant and equal to the phase of the bar $\Phi_{\rm{bar}}$. After this point, the phase $\Phi(R)$ starts varying with the radius. In this study, we consider the phase constant if its variation is $\Delta\Phi\leq\arcsin(0.15)$.

When a galaxy evolves in isolation, the major contribution to the phase of the second Fourier component in the disc density is the bar itself, since beyond its range the phase is averaged out.
However, in our cosmological simulations, the initial disc instabilities, the formation of a sporadic spiral arm, and the passage of some perturbers across the disc plane have, occasionally, the effect of significantly altering the total phase of the disc.
In order to account for this, and guide the analysis to the correct result, we do not compare $\Phi(R)$ with the phase of the whole disc \citep[as performed in][]{athanassoulaB}, but we substitute the total disc phase with the phase $\Phi(R_{\rm{max}})$ as our benchmark-phase, where $R_{\rm{max}}$ is the radius where $A_{2, \rm{max}}(<R)$ occurs (see Section \ref{ss:bar_stength}).

Although in the following we use $R_{\Phi}$ as the main bar length estimator (the values of $R_{\Phi}$ at $z=0$ are collected in Table~\ref{tab:summary_res}), it should be noted that $R_{\rm{max}}$, the radial coordinate where $A_{2}(<R)$ has its maximum, traces remarkably well the total extent of the stellar bar.
As a matter of fact, every annulus whose phase is coherent with $\Phi_{\rm{bar}}$ contributes to the increase of the cumulative function $A_{2}(<R)$. This results in the monotonic growth of $A_{2}(<R)$, which starts near the centre of the galaxy and ends as soon as this contribution is averaged out in the outer part of the disc and is no longer important (as it is exemplified in Figure~\ref{fig:examples}).

\begin{figure}
\includegraphics[width=\columnwidth]{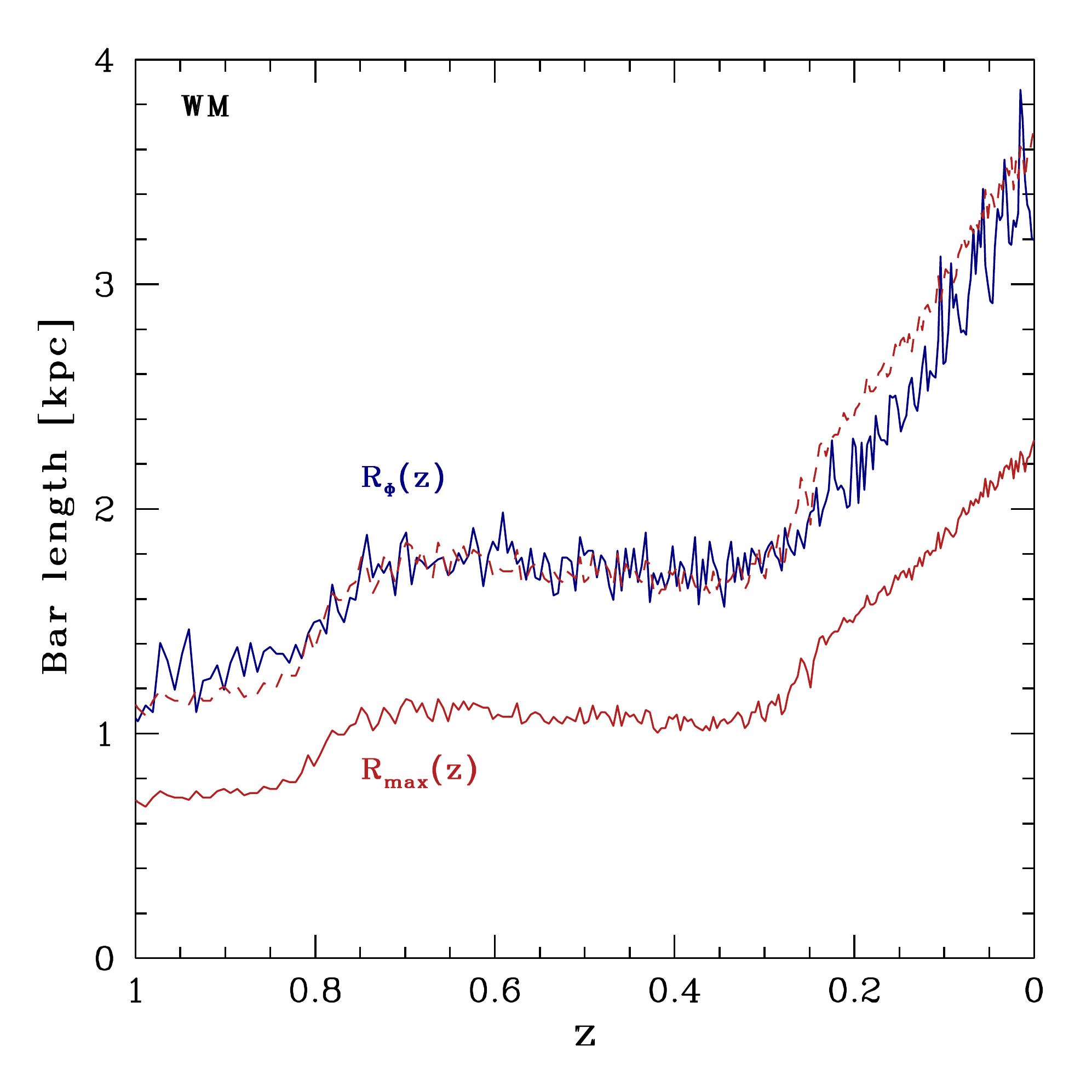}
\vspace{-0.5cm}
\caption{Extent estimators as a function of redshift: $R_{\rm{max}}(z)$ (red solid line), $\alpha R_{\rm{max}}(z)$ (red dashed line), and $R_{\Phi}(z)$ (blue line). The constant factor $\alpha=1.58$ (run WM) provides a remarkable agreement for $z<1$, i.e. the redshift range where a bar is clearly detectable [$A_{2,\rm{max}}(<R) \gsim 0.1$].}
\label{fig:agree}
\end{figure}

The results of these two criteria are plotted in Figure~\ref{fig:R} ($R_{\Phi}$ on the top and $R_{\rm{max}}$ on the bottom) adopting the same colour code used in Figure~\ref{fig:A2}.
In the plots, the black lines show the evolution of the inner non-axisymmetric component in run WM. They keep the same qualitative features of ErisBH (cyan) discussed in \cite{spinoso}.\footnote{The differences with respect to the analysis performed by \cite{spinoso} are due to the differences between ErisBH and run WM, as mentioned in section~\ref{subsec:ICsiblings}.}
The single peak of $R_{\rm{max}}$ at $z \approx 1.2$ is a tracer of the undergoing merger, since, in this method, the approaching satellite determines the position of the maximum asymmetry. This fluctuation is completely absent in the $R_{\Phi}$-method.
Aside from this small difference, both approaches agree with the general growth rate of the bar in all simulations.
We then quantitatively analyse the link between the two methods. In particular, for each run, we evaluated the parameter $\alpha$, in order to maximize the agreement between the function $\alpha R_{\rm{max}}(z)$ and $R_{\Phi}(z)$. The results (for $z \lsim 1$, where a clear bar is present) show that the parameters computed for the different simulation are very close to each other and their mean is equal to $<\!\alpha\!> = 1.61 \pm 0.17$.
As an example, the functions $R_{\rm{max}}(z)$ (solid red line), $R_{\Phi}(z)$ (solid blue line), and $\alpha R_{\rm{max}}(z)$ (red dashed line) for the run WM are plotted in Figure~\ref{fig:agree}.

It should be noted that the agreement factor $\alpha$ does (slightly) depend on the tolerance parameter $\Delta\Phi$.
The value we chose for $\Delta\Phi$ is motivated by a number of tests in which we tried to select only the real body of the bars, whilst avoiding phase fluctuations. The low standard deviation, proving that the constant $\alpha$ does not depend on redshift and on the specific run, demonstrates the affinity between the methods.  

\begin{figure*}
\begin{overpic}[width=\textwidth]{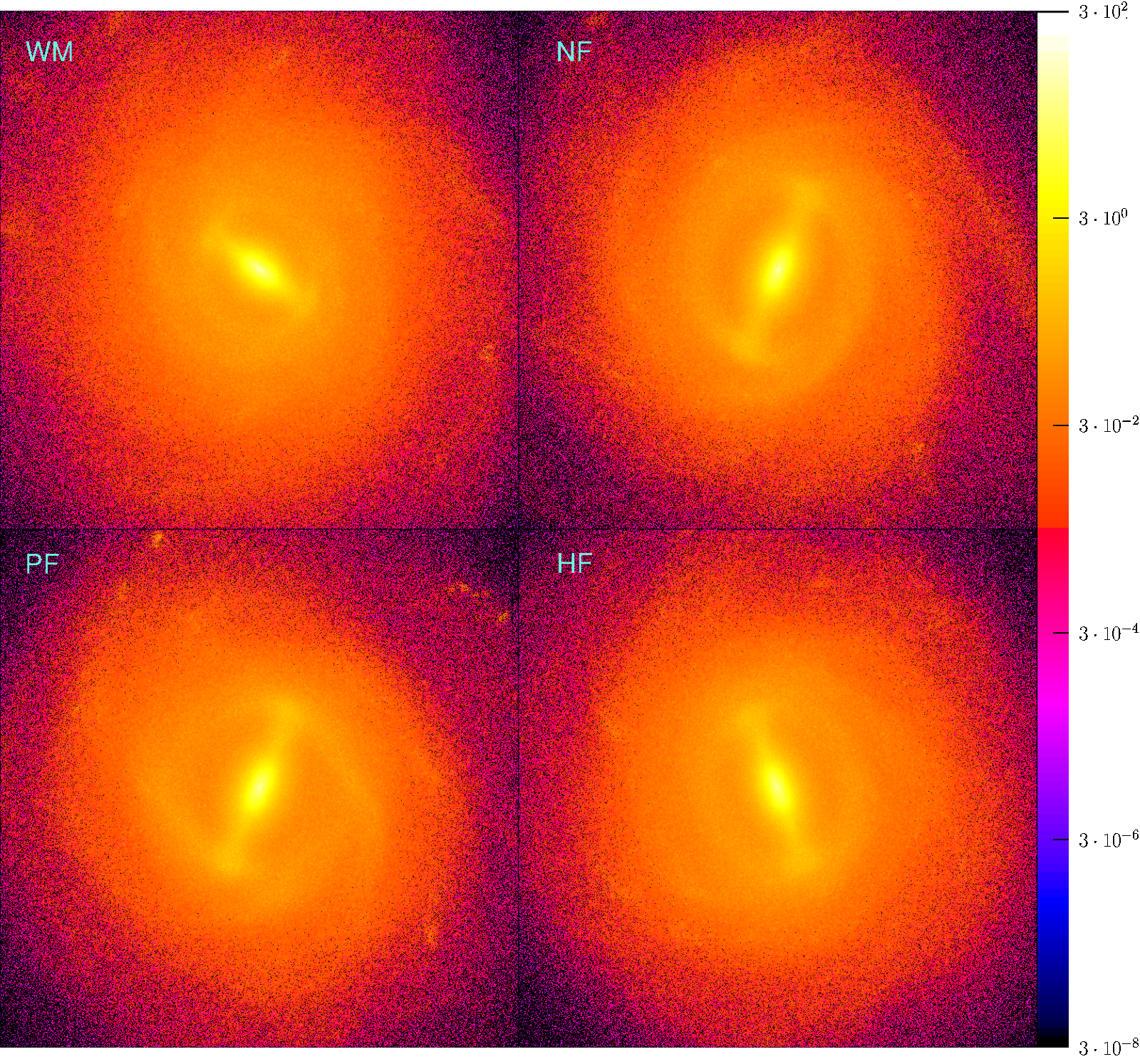}
\put(482,235){\colorbox{white}{\parbox{0.9cm}
	{
	\rotatebox{90}{$\left[\msun \cdot \rm{pc}^{-3}\right]$}
    }}}
\end{overpic}
\caption{Starting from the upper-left panel, face-on stellar density maps of runs WM, NF, PF, and HF at $z=0$. The colour code is the same used in Figure~\ref{fig:examples}, where also the map of run NM is provided.}
\label{fig:finalbars}
\end{figure*}

A transient period of bar weakening in run WM for $0.4 \lsim z \lsim 0.7$ is observed both in Figures~\ref{fig:A2} and \ref{fig:R}, and is possibly linked to the first pericentre of the fly-by. Such behaviour is consistent with a scenario in which minor mergers trigger the formation of the bar, whose growth, however, may be delayed by the fly-by-induced perturbation. 
To test such conclusion, we analyse the other runs, where the minor merger is absent. In such runs, despite the lack of the minor merger, the main galaxy develops a strong bar anyway, whose final length varies from $\gsim 4.3$ kpc (run NM) to $\gsim 5.1$ kpc (run NF). These parameters are summarised in Table~\ref{tab:summary_res}.
Figure~\ref{fig:finalbars} offers a qualitative comparison amongst the fully developed structures in each run at $z=0$ (note that the stellar density map of the run NM at $z=0$ is provided in Figure \ref{fig:examples} with the same colour code and scale). It is clear how the bar formation process does not depend on the different dynamical histories of the host galaxy.

The fly-by seems not to have any major impact on to the primary galaxy, either. While the first fly-by pericentre could be possibly associated to the onset of the bar growth ($z\approx 0.7$ in run NM), a strong bar is nevertheless already in place at the same redshift in the PF run, where the fly-by is still too far ($\sim 700$ kpc) to significantly perturb the galaxy. A similar evolution is observable in the HF run, in which the bar starts developing well before the first close passage. The fly-by effect here is only a possible delay of the growth of the bar, when the latter is 
already well formed.

\subsection{Properties of the evolved disc}

\subsubsection{Stellar density profile}


To check whether the bar evolution produces appreciable changes in the disc profile, we fit the stellar surface density maps of every galaxy at $z=0$ using a superposition of two S\'ersic profiles \citep{sersic}. The indices of the profile (listed in Table~\ref{tab:summary_res}) result close to unity for all the cases, typical for barred discs.

\begin{figure}
\includegraphics[width=\columnwidth]{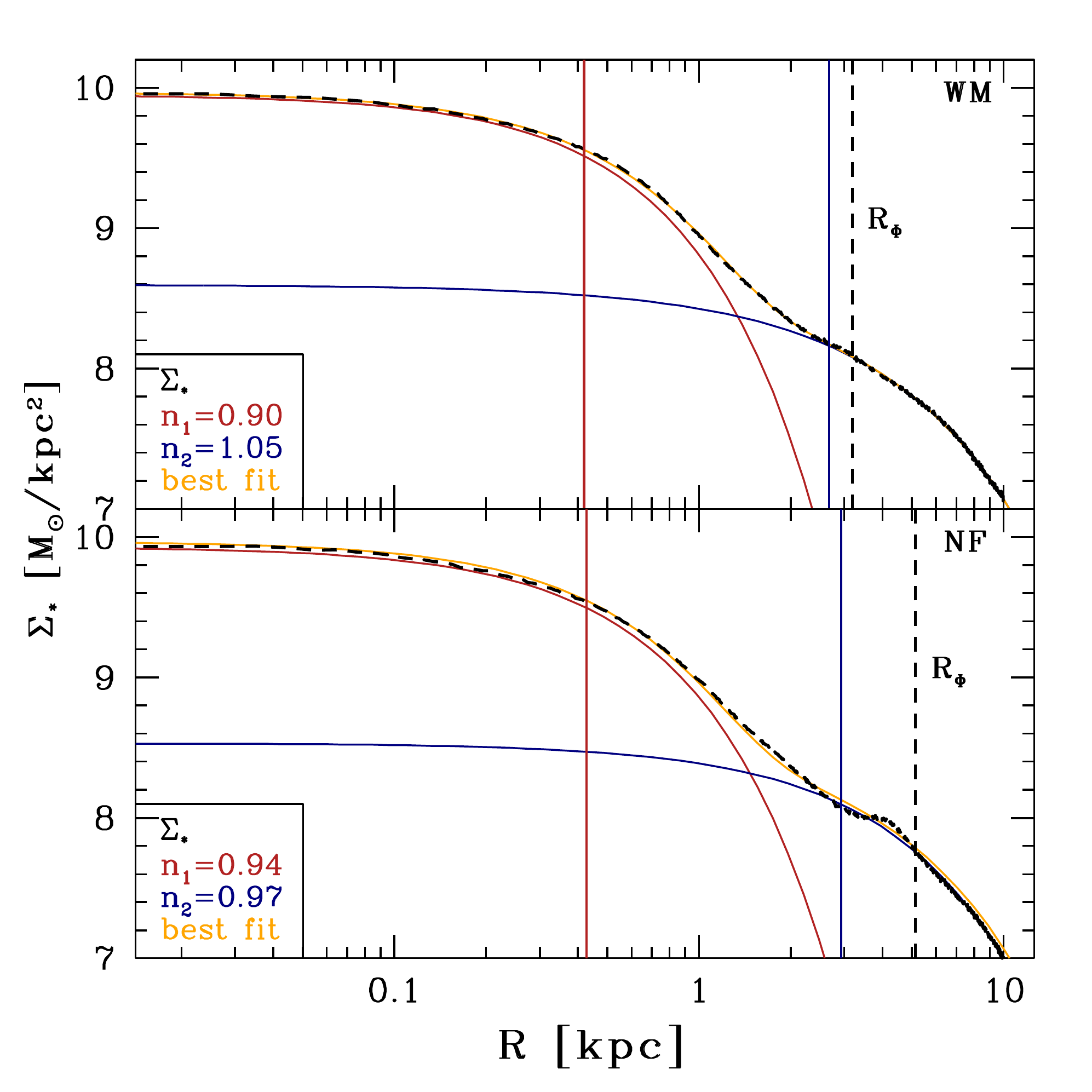}
\vspace{-0.5cm}
\caption{Profile decomposition of the stellar surface density for the WM (upper panel) and NF (lower panel) runs at $z=0$. The black dashed lines display the radial profile of the galactic disc as it has been calculated by integrating the mass density in cylindrical bin of height $8$ physical kpc. The yellow solid line shows the trend of the best fit, whereas the red and the blue lines refer to the inner and outer disc regions, respectively. The vertical solid lines mark the position of the scale radii of the corresponding fits with the same colour code. The black dashed lines show the bar length ($R_{\Phi}$).}
\label{fig:fit}
\end{figure}

Figure~\ref{fig:fit} shows the stellar surface distribution (black dashed line) as a function of the radius for the runs WM and NF which host two bars at different times of their development (bar growth in run NF has met fewer hurdles and appears to be stronger and longer). The yellow lines, which are the sum of the two S\'ersic profiles, show notable agreement with the surface density profile, whereas the red and blue lines follow the distribution of the inner and outer disc components, respectively. The vertical solid lines show the positions of the computed scale radii (red lines refer to the inner disc; blue lines to the outer disc). The black dashed lines mark the position of $R_{\Phi}$ at $z=0$, to provide a useful basis for comparison. We cannot state that there is an evident correlation between the stage of evolution of the bar (assessed by means of the strength and length estimators) and the positions of the scale radii just by assuming the five cases here analysed. In particular, bars in runs NM, NF, PF, and HF have similar evolutions at lower redshifts ($z \lsim 0.25$) and produce similar influences over the disc material. However, run WM, having the less evolved bar in our sample (see Figures~\ref{fig:A2} and \ref{fig:R}), shows the smallest scale radii and this is in agreement with e.g. \cite{athanassoulaB} and \cite{valenzuela}. It should be noted that the difference in the disc scale radius and extent could be linked to the mass accretion episode at $z \approx 1.2$.

\begin{figure}
\includegraphics[width=\columnwidth]{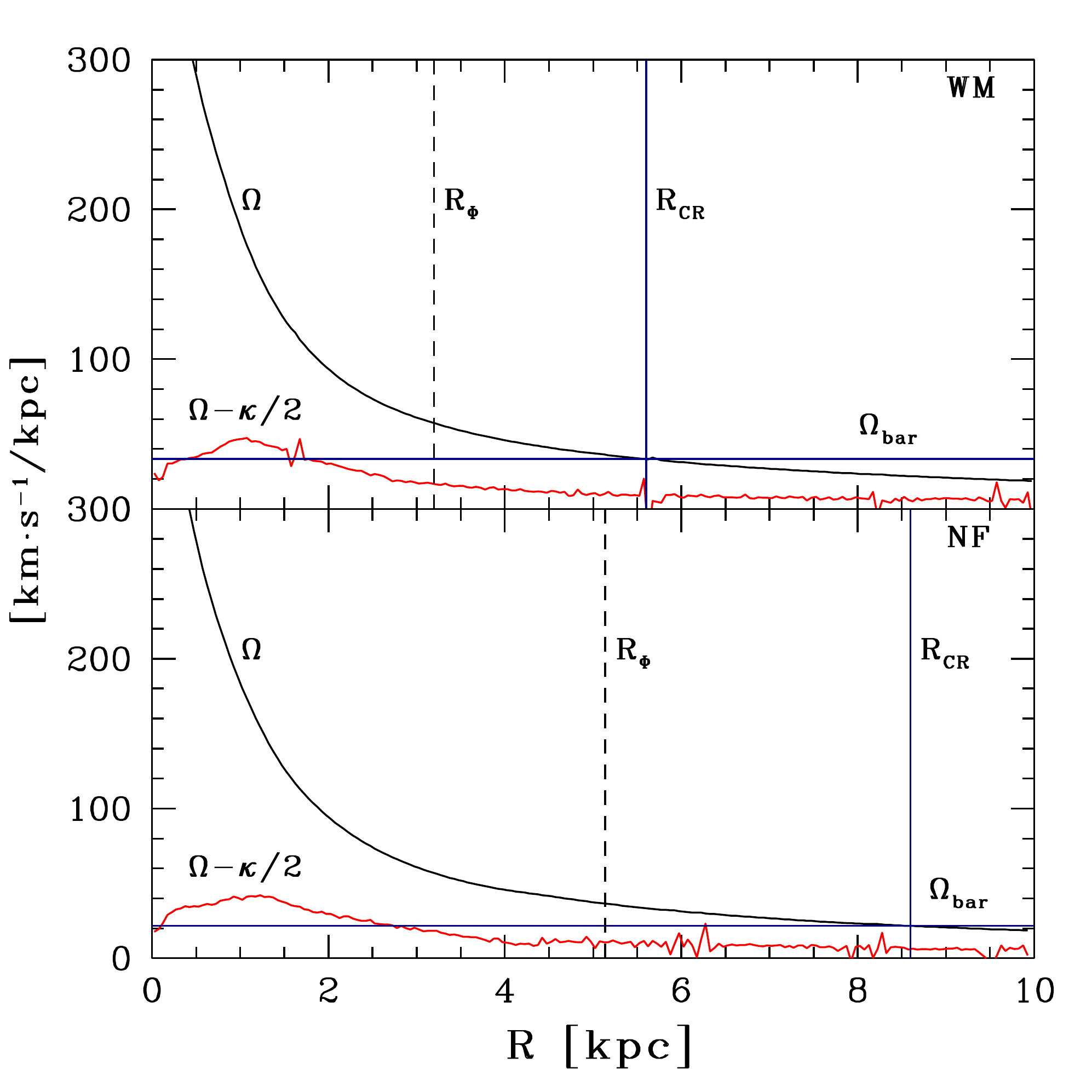}
\vspace{-0.5cm}
\caption{Frequency maps of the dominant galaxies in runs WM (upper panel) and NF (lower panel). As in Figure~\ref{fig:fit}, the black dashed lines refer to the bar extent at $z=0$. The black and red curves refer to the angular velocity $\Omega(R)$ and to the precession frequency $\Omega(R)-\kappa(R)/2$, respectively. The horizontal blue lines set the value of the bar angular velocities $\Omega_{\rm{bar}}$, evaluated at the end of their evolution, whereas the vertical lines highlight the positions of the corotational radii CR.}
\label{fig:frequency}
\end{figure}

\subsubsection{Frequency maps}


A useful aspect of our analysis is the study of the orbital frequencies of the fully evolved galaxies ($z=0$). 
Angular velocities and precession frequencies depend only on the gravitational potential and describe its inherent suitability to grow and sustain a bar.

Since we are dealing with bar-like structures, we are particularly interested in the precession frequency $\Omega-\kappa/2$, which defines those orbits which would close exactly after two radial oscillations over a period, in a frame corotating with the bar. Here, $\kappa$ is the epicycle frequency, which is linked to the angular velocity through the equation:
\begin{equation}
\kappa^2(R)=\left(R\frac{\rm{d}\Omega^2}{\rm{d}R}+4\Omega^2\right)_{R},
\label{eq:epicycle}
\end{equation}
where $\Omega$ is the angular velocity and $R$ the radius in cylindrical coordinates.
Equation~\ref{eq:epicycle} is valid in the epicycle approximation regime, where the extent of the radial oscillations of a point mass are considered negligible with respect to the radius \citep[for further details, see][]{binney}.

In Figure~\ref{fig:frequency}, we include the frequency maps for the same two cases (runs WM and NF in the upper and lower panel, respectively) shown in Figure~\ref{fig:fit}. In each panel, the black curve shows the angular velocity $\Omega(R)$ as a function of radius, whereas the red curve describes the trend of the two-fold precession curve $\Omega(R)-\kappa(R)/2$. The vertical blue line highlights the intersection between the bar frequency ($\Omega_{\rm{bar}}$, marked in the figure by the horizontal blue line) and the curve $\Omega(R)$. This point, the corotational radius $R_{\rm{CR}}$, has a major role in orbital theory and sets the upper limit for the bar extent according to \cite{contopoulos}.
The shape of the $\Omega-\kappa/2$ curve is typical for galaxies with low central concentration, as in our case. In both panels of Figure~\ref{fig:frequency}, $\Omega_{\rm{bar}}$ crosses the red curve in two points, since the frequency of the bar is below the peak of $\Omega-\kappa/2$: for these values of $R$ the galaxies have their Inner Lindblad Resonances.\footnote{The runs NM, PF, and HF (not shown in Figure~\ref{fig:frequency}) all have a similar behaviour.}

During the whole galaxy evolution, the bar absorbs angular momentum from the inner regions and redistributes it to the outer disc and the halo \citep{lynden, fuchs}. If the disc is not tidally perturbed, in this process, the bar usually slows down and increases its extent.
As a direct consequence of this frequency analysis, it is possible to identify the rate of bar rotation through the dimensionless parameter $\mathcal{R} = R_{\rm{CR}}/R_{\rm{bar}}$, where $R_{\rm{bar}}$ is the bar length (defined as $R_{\Phi}$ in this work).
As reported by \cite{debattistaB}, bars with $1.0 < \mathcal{R} < 1.4$ are called fast, whereas bars with $\mathcal{R} > 1.4$ are called slow.
The ratio $\mathcal{R}$ at $z=0$ is listed in Table~\ref{tab:summary_res} for each run. It is clear that, at the final stages of its evolution, each bar can be easily identified as slow, even if the dispersion of the values is large, emphasizing the different evolution histories of the various runs.

\subsubsection{B/P bulge emergence}

\begin{figure*}
\includegraphics[height=4.53cm]{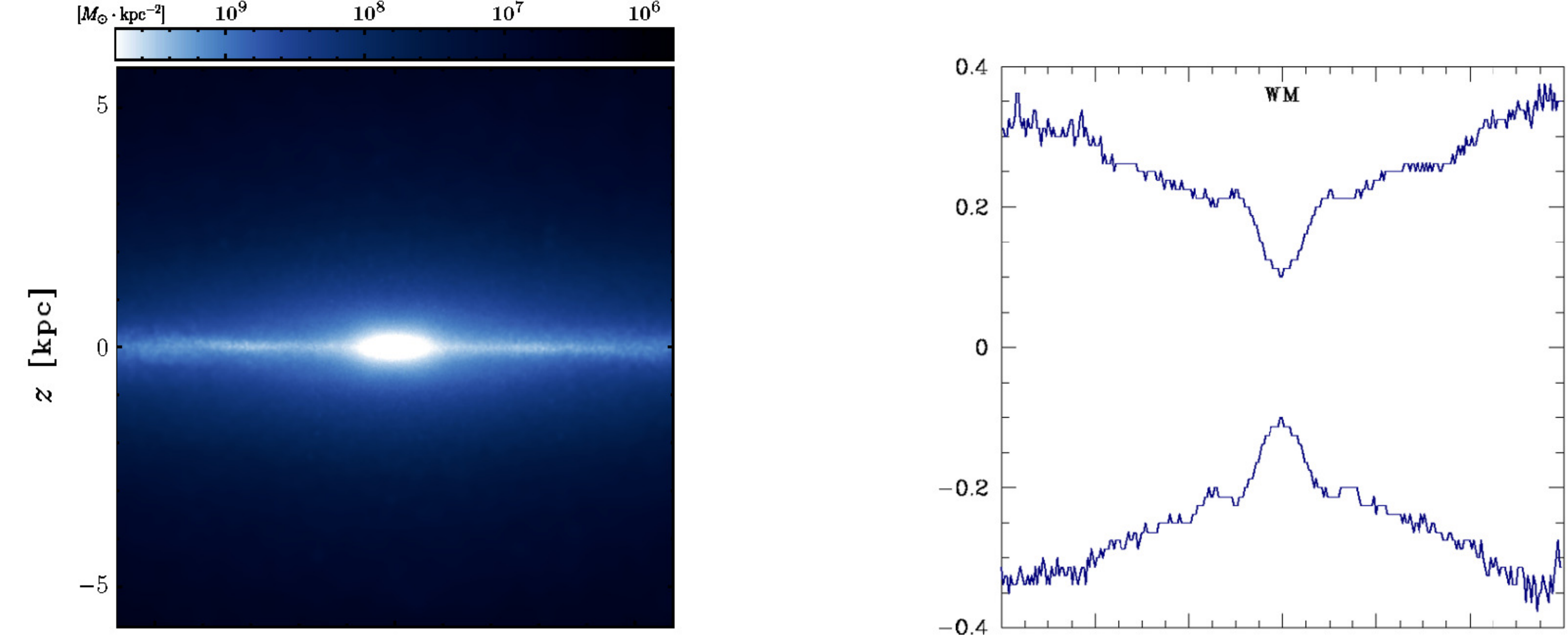}
\includegraphics[height=4.15cm]{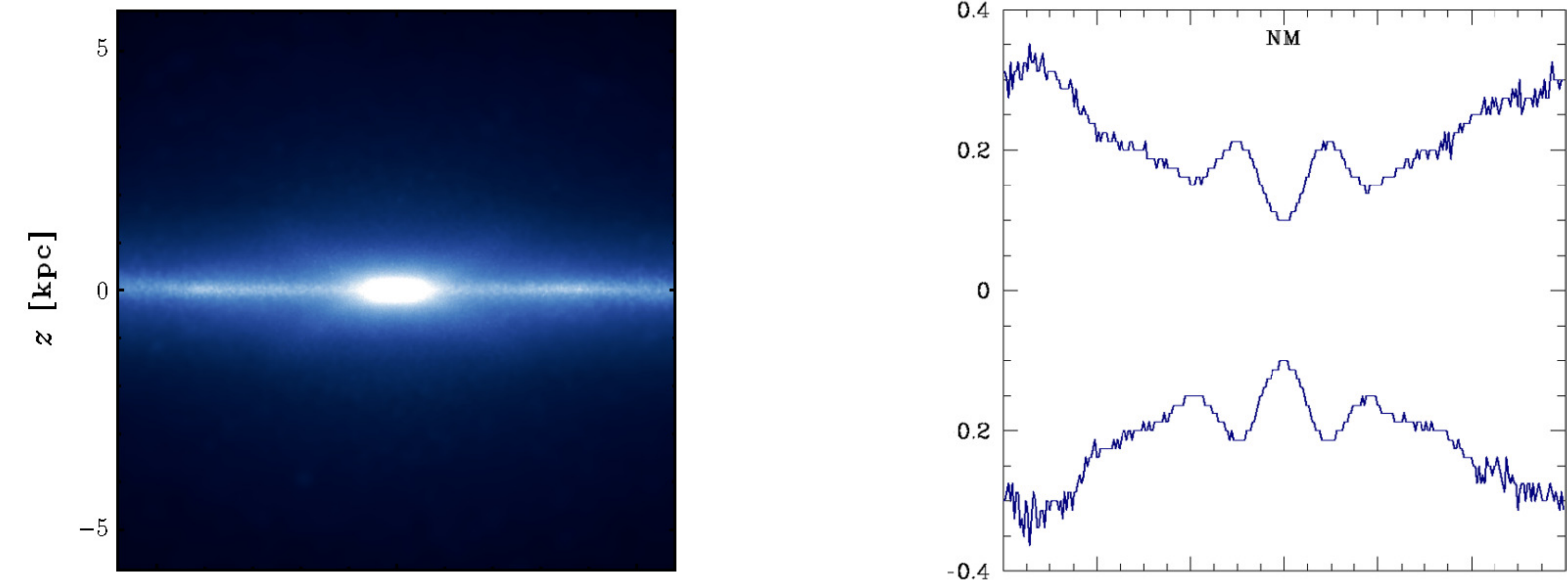}
\includegraphics[height=4.15cm]{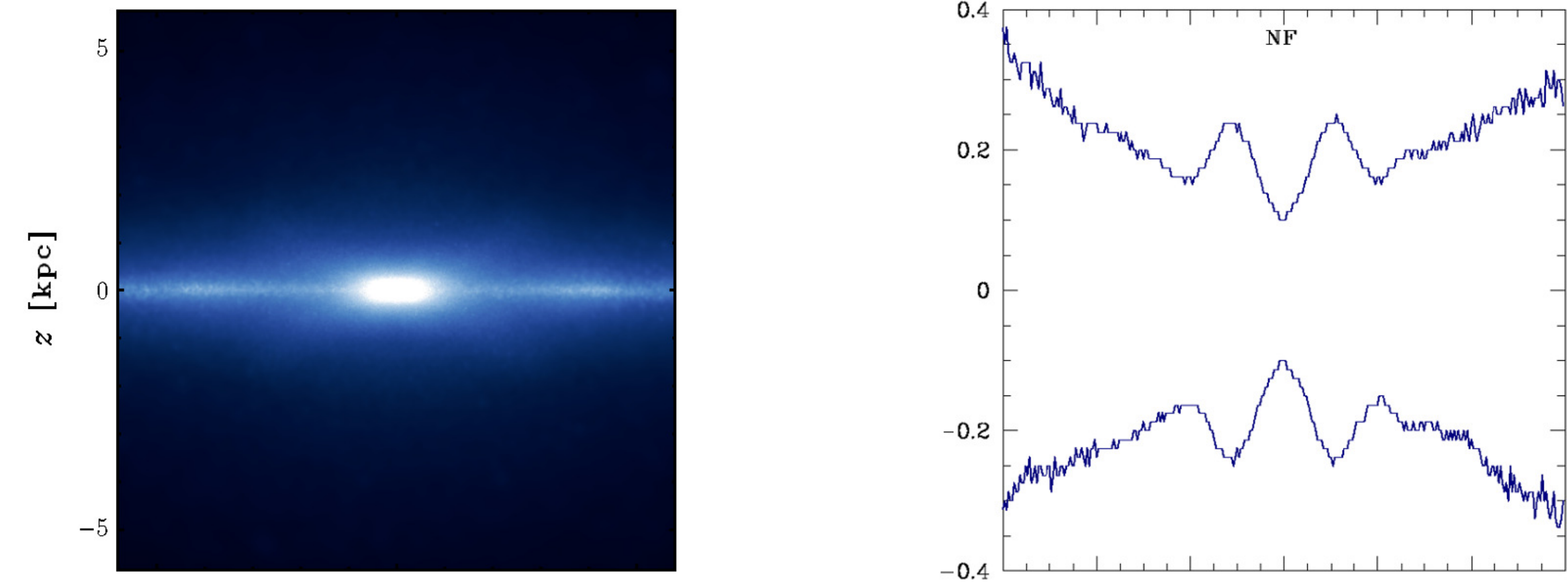}
\includegraphics[height=4.15cm]{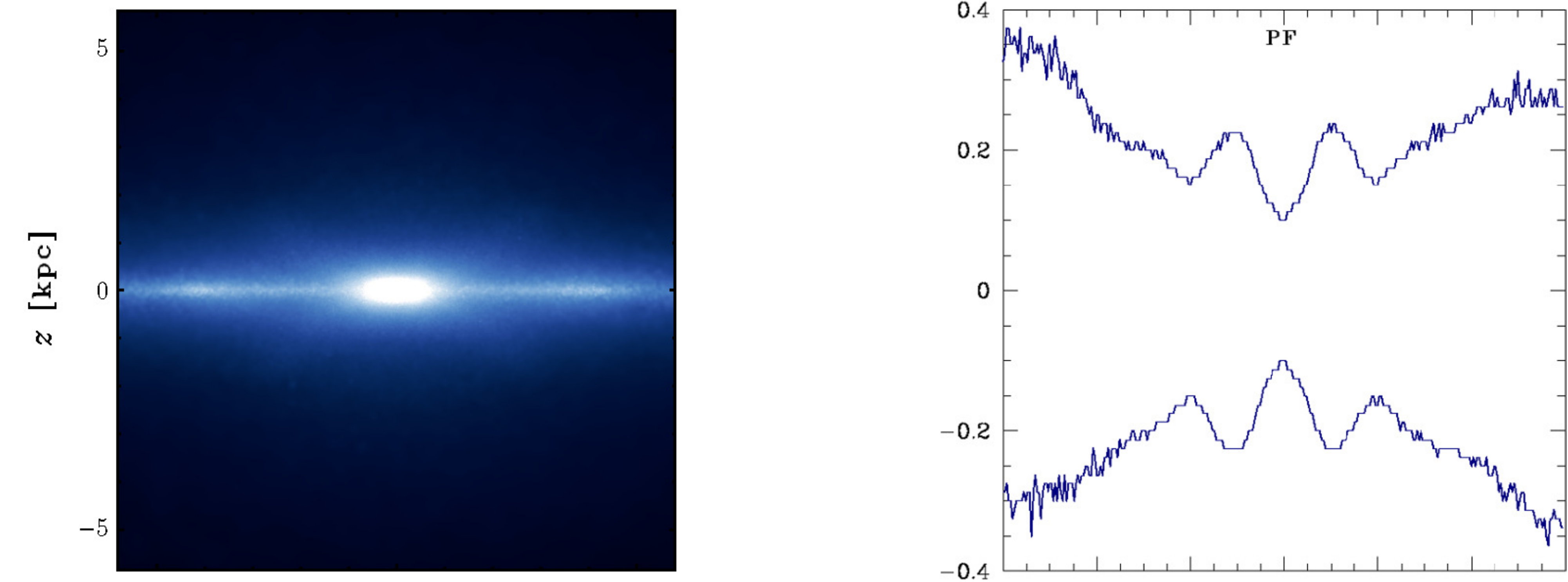}
\hspace*{1.78px}\includegraphics[height=4.67cm]{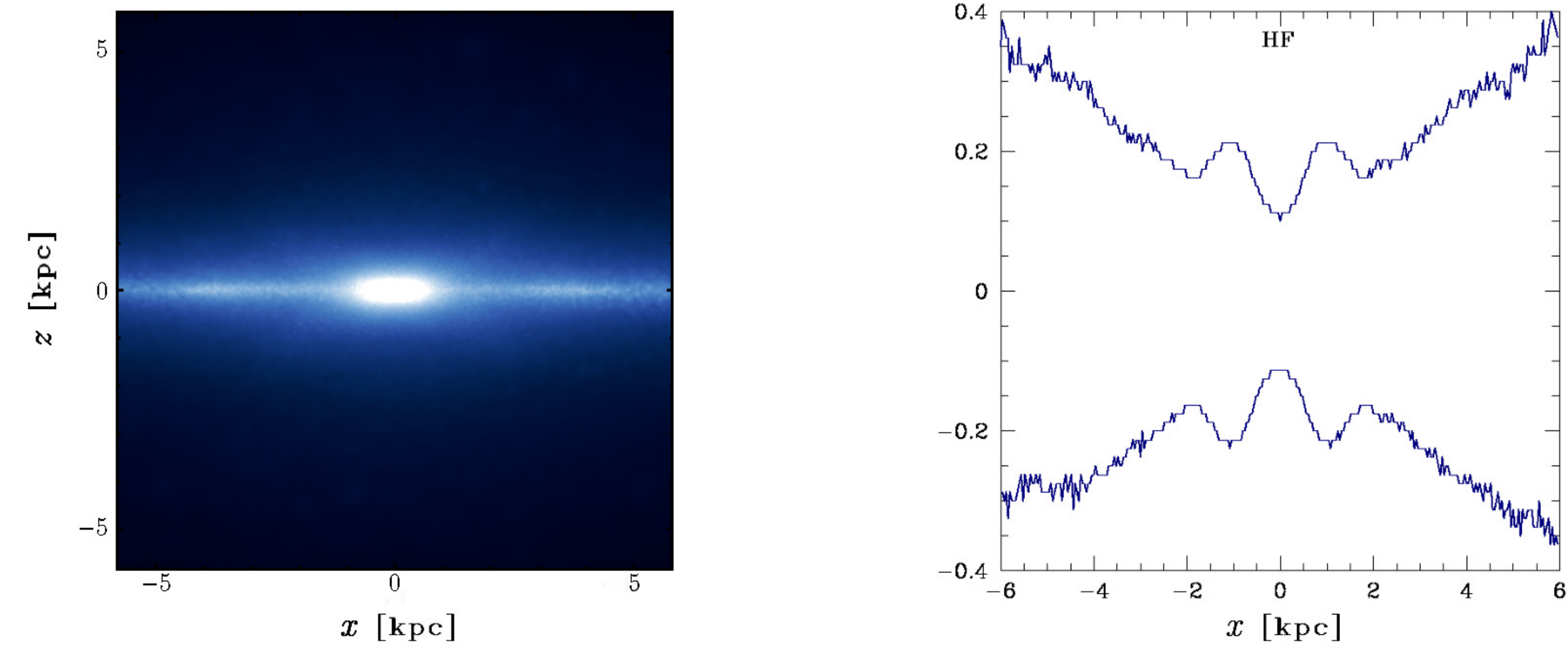}
\caption{On the left, the stellar surface density of the central galaxy is displayed edge-on with the bar major axis parallel to the $x$-axis. The colour code is logarithmically spaced and the units are $\msun/\rm{kpc}^2$. On the right, we show the position of the median density as a function of the distance from the centre of the galaxy, both for the region above ($z>0$) and below ($z<0$) the mid-plane. The scale on the $x$-axis is the same for the maps and the plots, whereas the $y$-axis scale is enlarged in the right-hand panels in order to highlight the variations. A developed X-shaped overdensity is immediately recognisable in every run, except in the case of run WM.}
\label{fig:boxyness}
\end{figure*}

Bars arise from the stellar disc and, therefore, at the beginning, they share with it the same vertical density profile. Despite this, after a few periods, bars experience some vertical instability processes when their growth continues without interference. This dynamical evolution is considered the cause of the so-called B/P bulges.
In order to assess its development in the various runs, we first select the edge-on projected density field with the bar aligned on the $x$-axis, for each run at $z=0$. We then subdivide the surface density field in rectangular bins on the $x$-axis (in this work we integrate the density on the $y$-axis over $15$ physical kpc), and evaluate the $z$-position of the median density in each bin in the positive and negative $z$-directions separately \citep[a similar approach has been used in][]{iannuzzi}.
The $z$-position of the median density as a function of the distance from the centre of the galaxy are shown on the right of Figure~\ref{fig:boxyness} for each run performed. In addition, each plot is accompanied by the corresponding surface density map, as it has been selected in the procedure described above.
The double-horned shape is clearly observable in almost all the cases, with the only exception of run WM. The clearest structure is visible in run NF.
As mentioned above, the bar in run WM is the least developed in our sample. On the contrary, the lack of significant perturbers in run NF favours the monotonic growth of the structure. For this reason, in run NF, vertical instabilities take place at earlier stages compared to run WM, resulting in a more evolved X-shape.
According to this scenario, the bar in run WM would face a similar fate in its future evolution.

\begin{table}
\centering
\caption{Summary of the most relevant properties of the final ($z=0$) bar for each run. From left to right, columns contain the run identifier, the bar extent $R_{\Phi}$, the bar length in units of disc scale-radius, the ratio $R_{\rm{CR}}/R_{\rm{bar}}$, the indices of the S\'ersic profiles adopted to fit the disc surface density, and the presence of the B/P structure.}
\label{tab:summary_res}
\begin{tabular}{cccccccc}
	\hline	
	Run & $R_{\Phi}$ & $R_{\Phi}/R_{\rm disc}$ & $\mathcal{R}$ & $n_{1}$ & $n_{2}$ & B/P \\
	\hline \hline
	WM & 3.195 & 1.194 & 1.752739 & 0.90 & 1.05 & no\\
	\hline
	NM & 4.355 & 1.274 & 1.595867 & 0.94 & 0.86 & yes\\
	\hline		
	NF & 5.135 & 1.753 & 1.674781 & 0.94 & 0.97 & yes\\
	\hline		
	PF & 4.485 & 1.642 & 1.884058 & 0.93 & 0.99 & yes\\
	\hline
	HF & 4.615 & 2.045 & 1.516793 & 0.84 & 1.11 & yes\\
	\hline
\end{tabular}
\end{table}

\section{Discussion and Conclusions}
\label{sec:conclusion}

In our work, we assessed the nature of the bar resulting from a fully consistent cosmological simulation of a Milky Way-sized galaxy. In particular, we analysed its late evolutionary stages ($z\lsim 2$), when the central regions are already prone to bar instability, and we checked whether the seed perturbation that triggers the actual bar growth is due to secular/internal processes only, or to tidal perturbations exerted by the evolving background of satellite galaxies.

In our study, we ran multiple and different versions of the final phases of the ErisBH simulation, allowing for a modification of the sequence of interactions amongst the main galaxy and its satellites. More specifically, we removed the only minor merger occurring in the galaxy evolutionary history after $z \approx 1.8$, and we changed either the orbital or structural parameters of the closest fly-by in the system, or we simply avoided its formation in the galactic neighbourhood. Although all the different runs start at the same point and with the same exact physics and sub-grid physics modules, we do perturb the system to some extent by removing or changing the location of a subset of particles, i.e. modifying the energy and mass content in the box. 
 
We demonstrated that the origin of the bar forming in ErisBH run is not linked to any particular tidal event. Indeed, an even stronger and longer bar forms when the merger does not occur at all.
Furthermore, the B/P shape (see Figure~\ref{fig:boxyness}) reflects the bar evolutionary stage, which is more advanced in the runs NM, NF, PF, and HF with respect to run WM.
Analogously, the outcome of our experiments shows that we have not introduced any relevant numerical artefact with our procedure as this would have had an impact on the dynamical evolution and on the development of the bar itself. This last point is also evident in Figure~\ref{fig:mass}, where the primary galaxy evolves almost identically in all cases, suggesting that the numerical noise introduced by the engineering of the dynamical state of the system does not lead to any numerical instability.

Moreover, it seems that the fly-by could delay the growth of a bar if such structure is already present, but certainly this interaction cannot be considered itself the trigger of the bar growth, and does not alter the further evolution of the bar properties. The delay could be due to the increase in the stellar velocity dispersion imprinted by the fly-by, resulting in a more stable disc \citep{guedesB}.

We further comment that another run was performed with different initial conditions. This case is similar to the case of HF, but the dark matter component of the fly-by has been multiplied five times. We decided not to discuss it here, since its merger history (a further merger takes place at $z \lsim 1$) does produce a significantly different object at $z=0$ and, therefore, it cannot be directly compared with the other siblings.
Notwithstanding this, the case would fall into the category of the objects whose bar has been strongly suppressed.
As mentioned above, some studies claim that the global effect of the gravitational perturbers could be summarised in a growth/formation inhibition \citep{li, lin}.

It is interesting to compare our results to similar recent findings obtained by \cite{moetazedian}, who detailed the effect of low-mass satellites on the bar growth in a bar-unstable galaxy. 
The masses, tidal radii, and impact parameters of the perturbers are inferred from the cosmological Aquarius-D simulation \citep{springelAqua}. In agreement with our investigation, \cite{moetazedian} find that the only effect of the satellites is to either slightly anticipate or delay the bar growth (by at most $\sim$ 1 Gyr), whereas the main properties of the bar as well as its growth rate remain almost identical. We stress that our results and those discussed in \cite{moetazedian} complement each other. On one hand, our study takes into account the full assembly history of the main galaxy as well as the effect of gas, whereas both these aspects are not included in \cite{moetazedian}. On the other hand, the idealised nature of the \cite{moetazedian} initial conditions results in an higher level of axisymmetry, which allows for a better characterization of the initial seeding of the bar. As an example, \cite{moetazedian} were able to 
follow the evolution of the bar strength parameter $A_{2}$\footnote{Note that the parameter defined as $A_{2}$ in this work is equivalent to $A_{2}/A_{0}$ in \cite{moetazedian}.} down to fluctuations of the order of $\simeq 10^{-3}$, well below the physical noise due to the small substructures present in our study. The consistency of the findings in such different studies strongly supports the internal/secular origin of bars in field galaxies. 

In summary, we proved that there is no need of a major external perturbation for the formation of the bar in ErisBH. As long as the galactic potential is prone to bar instability, as in the last stages of ErisBH \citep{spinoso}, a bar develops independently of any interaction (if these interactions are not strong enough to considerably modify the galactic potential). This points to the self-gravity of the disc and to its interplay with various internal processes, from cooling to energy input by SN feedback, as the main driver of bar formation and growth, similarly to what seen in simulations of secular evolution of massive spirals \citep{debattista}.

At high redshifts, when galactic discs are still low in mass (hence weakly self-gravitating) and perturbations by tidal interactions and mergers with massive satellites are much more frequent, the role of tidal triggering in bar formation and growth might be more important, as suggested by \cite{guedesB} and \cite{fiacconi}. We argue here that massive spiral galaxies at $z\lsim 1$ should, instead, belong to the same regime we studied in this paper. This means that the standard scenario of secular evolution does hold, at low redshifts, even in the complex context of hierarchical galaxy formation.

\section*{Acknowledgements}
The authors would like to thank the anonymous reviewer for the helpful comments that improved the quality of the manuscript.
TZ thanks Thomas R. Quinn for his help and useful discussions.
We acknowledge the CINECA award under the ISCRA initiative, for the availability of high-performance computing resources and support. Numerical calculations have been complemented on Piz Daint, at the Swiss National Supercomputing Center (CSCS), and through a CINECA-INFN agreement, providing access to resources on GALILEO and MARCONI at CINECA. PRC acknowledges support by the Tomalla foundation.



\bibliographystyle{mnras}
\bibliography{bibliography}

\begin{thebibliography}{}
\makeatletter
\relax
\def\mn@urlcharsother{\let\do\@makeother \do\$\do\&\do\#\do\^\do\_\do\%\do\~}
\def\mn@doi{\begingroup\mn@urlcharsother \@ifnextchar [ {\mn@doi@}
  {\mn@doi@[]}}
\def\mn@doi@[#1]#2{\def\@tempa{#1}\ifx\@tempa\@empty \href
  {http://dx.doi.org/#2} {doi:#2}\else \href {http://dx.doi.org/#2} {#1}\fi
  \endgroup}
\def\mn@eprint#1#2{\mn@eprint@#1:#2::\@nil}
\def\mn@eprint@arXiv#1{\href {http://arxiv.org/abs/#1} {{\tt arXiv:#1}}}
\def\mn@eprint@dblp#1{\href {http://dblp.uni-trier.de/rec/bibtex/#1.xml}
  {dblp:#1}}
\def\mn@eprint@#1:#2:#3:#4\@nil{\def\@tempa {#1}\def\@tempb {#2}\def\@tempc
  {#3}\ifx \@tempc \@empty \let \@tempc \@tempb \let \@tempb \@tempa \fi \ifx
  \@tempb \@empty \def\@tempb {arXiv}\fi \@ifundefined
  {mn@eprint@\@tempb}{\@tempb:\@tempc}{\expandafter \expandafter \csname
  mn@eprint@\@tempb\endcsname \expandafter{\@tempc}}}

\bibitem[\protect\citeauthoryear{{Algorry} et~al.,}{{Algorry}
  et~al.}{2017}]{algorry}
{Algorry} D.~G.,  et~al., 2017, \mn@doi [\mnras] {10.1093/mnras/stx1008}, \href
  {http://adsabs.harvard.edu/abs/2017MNRAS.469.1054A} {469, 1054}

\bibitem[\protect\citeauthoryear{{Athanassoula}}{{Athanassoula}}{1992}]{athanassoula}
{Athanassoula} E.,  1992, \mn@doi [\mnras] {10.1093/mnras/259.2.345}, \href
  {http://adsabs.harvard.edu/abs/1992MNRAS.259..345A} {259, 345}

\bibitem[\protect\citeauthoryear{{Athanassoula}}{{Athanassoula}}{2008}]{athanassoulaC}
{Athanassoula} E.,  2008, in {Bureau} M.,  {Athanassoula} E.,   {Barbuy} B.,
  eds,  IAU Symposium Vol. 245, Formation and Evolution of Galaxy Bulges. pp
  93--102, \mn@doi{10.1017/S1743921308017389}

\bibitem[\protect\citeauthoryear{{Athanassoula} \& {Misiriotis}}{{Athanassoula}
  \& {Misiriotis}}{2002}]{athanassoulaB}
{Athanassoula} E.,  {Misiriotis} A.,  2002, \mn@doi [\mnras]
  {10.1046/j.1365-8711.2002.05028.x}, \href
  {http://adsabs.harvard.edu/abs/2002MNRAS.330...35A} {330, 35}

\bibitem[\protect\citeauthoryear{{Bellovary}, {Governato}, {Quinn}, {Wadsley},
  {Shen}  \& {Volonteri}}{{Bellovary} et~al.}{2010}]{bellovary}
{Bellovary} J.~M.,  {Governato} F.,  {Quinn} T.~R.,  {Wadsley} J.,  {Shen} S.,
   {Volonteri} M.,  2010, \mn@doi [\apjl] {10.1088/2041-8205/721/2/L148}, \href
  {http://adsabs.harvard.edu/abs/2010ApJ...721L.148B} {721, L148}

\bibitem[\protect\citeauthoryear{{Binney} \& {Tremaine}}{{Binney} \&
  {Tremaine}}{2008}]{binney}
{Binney} J.,  {Tremaine} S.,  2008, {Galactic Dynamics: Second Edition}.
Princeton University Press

\bibitem[\protect\citeauthoryear{{Bondi}}{{Bondi}}{1952}]{bondiB}
{Bondi} H.,  1952, \mn@doi [\mnras] {10.1093/mnras/112.2.195}, \href
  {http://adsabs.harvard.edu/abs/1952MNRAS.112..195B} {112, 195}

\bibitem[\protect\citeauthoryear{{Bondi} \& {Hoyle}}{{Bondi} \&
  {Hoyle}}{1944}]{bondi}
{Bondi} H.,  {Hoyle} F.,  1944, \mn@doi [\mnras] {10.1093/mnras/104.5.273},
  \href {http://adsabs.harvard.edu/abs/1944MNRAS.104..273B} {104, 273}

\bibitem[\protect\citeauthoryear{{Bonoli}, {Mayer}, {Kazantzidis}, {Madau},
  {Bellovary}  \& {Governato}}{{Bonoli} et~al.}{2016}]{bonoli}
{Bonoli} S.,  {Mayer} L.,  {Kazantzidis} S.,  {Madau} P.,  {Bellovary} J.,
  {Governato} F.,  2016, \mn@doi [\mnras] {10.1093/mnras/stw694}, \href
  {http://adsabs.harvard.edu/abs/2016MNRAS.459.2603B} {459, 2603}

\bibitem[\protect\citeauthoryear{{Bureau} \& {Athanassoula}}{{Bureau} \&
  {Athanassoula}}{2005}]{bureau}
{Bureau} M.,  {Athanassoula} E.,  2005, \mn@doi [\apj] {10.1086/430056}, \href
  {http://adsabs.harvard.edu/abs/2005ApJ...626..159B} {626, 159}

\bibitem[\protect\citeauthoryear{{Byrd}, {Valtonen}, {Valtaoja}  \&
  {Sundelius}}{{Byrd} et~al.}{1986}]{byrd}
{Byrd} G.~G.,  {Valtonen} M.~J.,  {Valtaoja} L.,   {Sundelius} B.,  1986, \aap,
  \href {http://adsabs.harvard.edu/abs/1986A\%26A...166...75B} {166, 75}

\bibitem[\protect\citeauthoryear{{Cheung} et~al.,}{{Cheung}
  et~al.}{2013}]{cheung}
{Cheung} E.,  et~al., 2013, \mn@doi [\apj] {10.1088/0004-637X/779/2/162}, \href
  {http://adsabs.harvard.edu/abs/2013ApJ...779..162C} {779, 162}

\bibitem[\protect\citeauthoryear{{Combes}, {Debbasch}, {Friedli}  \&
  {Pfenniger}}{{Combes} et~al.}{1990}]{combes}
{Combes} F.,  {Debbasch} F.,  {Friedli} D.,   {Pfenniger} D.,  1990, \aap,
  \href {http://adsabs.harvard.edu/abs/1990A\%26A...233...82C} {233, 82}

\bibitem[\protect\citeauthoryear{{Consolandi}, {Gavazzi}, {Fumagalli}, {Dotti}
  \& {Fossati}}{{Consolandi} et~al.}{2016}]{consolandiB}
{Consolandi} G.,  {Gavazzi} G.,  {Fumagalli} M.,  {Dotti} M.,   {Fossati} M.,
  2016, \mn@doi [\aap] {10.1051/0004-6361/201527618}, \href
  {http://adsabs.harvard.edu/abs/2016A\%26A...591A..38C} {591, A38}

\bibitem[\protect\citeauthoryear{{Consolandi}, {Dotti}, {Boselli}, {Gavazzi}
  \& {Gargiulo}}{{Consolandi} et~al.}{2017}]{consolandi}
{Consolandi} G.,  {Dotti} M.,  {Boselli} A.,  {Gavazzi} G.,   {Gargiulo} F.,
  2017, \mn@doi [\aap] {10.1051/0004-6361/201629213}, \href
  {http://adsabs.harvard.edu/abs/2017A\%26A...598A.114C} {598, A114}

\bibitem[\protect\citeauthoryear{{Contopoulos}}{{Contopoulos}}{1980}]{contopoulos}
{Contopoulos} G.,  1980, \aap, \href
  {http://adsabs.harvard.edu/abs/1980A\%26A....81..198C} {81, 198}

\bibitem[\protect\citeauthoryear{{Curir}, {Mazzei}  \& {Murante}}{{Curir}
  et~al.}{2006}]{curir}
{Curir} A.,  {Mazzei} P.,   {Murante} G.,  2006, \mn@doi [\aap]
  {10.1051/0004-6361:20053418}, \href
  {http://adsabs.harvard.edu/abs/2006A\%26A...447..453C} {447, 453}

\bibitem[\protect\citeauthoryear{{Debattista} \& {Sellwood}}{{Debattista} \&
  {Sellwood}}{2000}]{debattistaB}
{Debattista} V.~P.,  {Sellwood} J.~A.,  2000, \mn@doi [\apj] {10.1086/317148},
  \href {http://adsabs.harvard.edu/abs/2000ApJ...543..704D} {543, 704}

\bibitem[\protect\citeauthoryear{{Debattista}, {Mayer}, {Carollo}, {Moore},
  {Wadsley}  \& {Quinn}}{{Debattista} et~al.}{2006}]{debattista}
{Debattista} V.~P.,  {Mayer} L.,  {Carollo} C.~M.,  {Moore} B.,  {Wadsley} J.,
   {Quinn} T.,  2006, \mn@doi [\apj] {10.1086/504147}, \href
  {http://adsabs.harvard.edu/abs/2006ApJ...645..209D} {645, 209}

\bibitem[\protect\citeauthoryear{{Fanali}, {Dotti}, {Fiacconi}  \&
  {Haardt}}{{Fanali} et~al.}{2015}]{fanali}
{Fanali} R.,  {Dotti} M.,  {Fiacconi} D.,   {Haardt} F.,  2015, \mn@doi
  [\mnras] {10.1093/mnras/stv2247}, \href
  {http://adsabs.harvard.edu/abs/2015MNRAS.454.3641F} {454, 3641}

\bibitem[\protect\citeauthoryear{{Fiacconi}, {Feldmann}  \& {Mayer}}{{Fiacconi}
  et~al.}{2015}]{fiacconi}
{Fiacconi} D.,  {Feldmann} R.,   {Mayer} L.,  2015, \mn@doi [\mnras]
  {10.1093/mnras/stu2228}, \href
  {http://adsabs.harvard.edu/abs/2015MNRAS.446.1957F} {446, 1957}

\bibitem[\protect\citeauthoryear{{Fuchs} \& {Athanassoula}}{{Fuchs} \&
  {Athanassoula}}{2005}]{fuchs}
{Fuchs} B.,  {Athanassoula} E.,  2005, \mn@doi [\aap]
  {10.1051/0004-6361:20052894}, \href
  {http://adsabs.harvard.edu/abs/2005A\%26A...444..455F} {444, 455}

\bibitem[\protect\citeauthoryear{{Gauthier}, {Dubinski}  \&
  {Widrow}}{{Gauthier} et~al.}{2006}]{gauthier}
{Gauthier} J.-R.,  {Dubinski} J.,   {Widrow} L.~M.,  2006, \mn@doi [\apj]
  {10.1086/508860}, \href {http://adsabs.harvard.edu/abs/2006ApJ...653.1180G}
  {653, 1180}

\bibitem[\protect\citeauthoryear{{Gavazzi} et~al.,}{{Gavazzi}
  et~al.}{2015}]{gavazzi}
{Gavazzi} G.,  et~al., 2015, \mn@doi [\aap] {10.1051/0004-6361/201425351},
  \href {http://adsabs.harvard.edu/abs/2015A\%26A...580A.116G} {580, A116}

\bibitem[\protect\citeauthoryear{{Goz}, {Monaco}, {Murante}  \& {Curir}}{{Goz}
  et~al.}{2015}]{goz}
{Goz} D.,  {Monaco} P.,  {Murante} G.,   {Curir} A.,  2015, \mn@doi [\mnras]
  {10.1093/mnras/stu2557}, \href
  {http://adsabs.harvard.edu/abs/2015MNRAS.447.1774G} {447, 1774}

\bibitem[\protect\citeauthoryear{{Guedes}, {Callegari}, {Madau}  \&
  {Mayer}}{{Guedes} et~al.}{2011}]{guedes}
{Guedes} J.,  {Callegari} S.,  {Madau} P.,   {Mayer} L.,  2011, \mn@doi [\apj]
  {10.1088/0004-637X/742/2/76}, \href
  {http://adsabs.harvard.edu/abs/2011ApJ...742...76G} {742, 76}

\bibitem[\protect\citeauthoryear{{Guedes}, {Mayer}, {Carollo}  \&
  {Madau}}{{Guedes} et~al.}{2013}]{guedesB}
{Guedes} J.,  {Mayer} L.,  {Carollo} M.,   {Madau} P.,  2013, \mn@doi [\apj]
  {10.1088/0004-637X/772/1/36}, \href
  {http://adsabs.harvard.edu/abs/2013ApJ...772...36G} {772, 36}

\bibitem[\protect\citeauthoryear{{Haardt} \& {Madau}}{{Haardt} \&
  {Madau}}{1996}]{haardt}
{Haardt} F.,  {Madau} P.,  1996, \mn@doi [\apj] {10.1086/177035}, \href
  {http://adsabs.harvard.edu/abs/1996ApJ...461...20H} {461, 20}

\bibitem[\protect\citeauthoryear{{Hakobyan} et~al.,}{{Hakobyan}
  et~al.}{2016}]{hakobyan}
{Hakobyan} A.~A.,  et~al., 2016, \mn@doi [\mnras] {10.1093/mnras/stv2853},
  \href {http://adsabs.harvard.edu/abs/2016MNRAS.456.2848H} {456, 2848}

\bibitem[\protect\citeauthoryear{{Ho}, {Filippenko}  \& {Sargent}}{{Ho}
  et~al.}{1997}]{ho}
{Ho} L.~C.,  {Filippenko} A.~V.,   {Sargent} W.~L.~W.,  1997, \mn@doi [\apj]
  {10.1086/304643}, \href {http://adsabs.harvard.edu/abs/1997ApJ...487..591H}
  {487, 591}

\bibitem[\protect\citeauthoryear{{Hohl}}{{Hohl}}{1971}]{hohl}
{Hohl} F.,  1971, \mn@doi [\apj] {10.1086/151091}, \href
  {http://adsabs.harvard.edu/abs/1971ApJ...168..343H} {168, 343}

\bibitem[\protect\citeauthoryear{{Hoyle} \& {Lyttleton}}{{Hoyle} \&
  {Lyttleton}}{1939}]{hoyle}
{Hoyle} F.,  {Lyttleton} R.~A.,  1939, \mn@doi [Proceedings of the Cambridge
  Philosophical Society] {10.1017/S0305004100021150}, \href
  {http://adsabs.harvard.edu/abs/1939PCPS...35..405H} {35, 405}

\bibitem[\protect\citeauthoryear{{Hunt} \& {Malkan}}{{Hunt} \&
  {Malkan}}{1999}]{hunt}
{Hunt} L.~K.,  {Malkan} M.~A.,  1999, \mn@doi [\apj] {10.1086/307150}, \href
  {http://adsabs.harvard.edu/abs/1999ApJ...516..660H} {516, 660}

\bibitem[\protect\citeauthoryear{{Iannuzzi} \& {Athanassoula}}{{Iannuzzi} \&
  {Athanassoula}}{2015}]{iannuzzi}
{Iannuzzi} F.,  {Athanassoula} E.,  2015, \mn@doi [\mnras]
  {10.1093/mnras/stv764}, \href
  {http://adsabs.harvard.edu/abs/2015MNRAS.450.2514I} {450, 2514}

\bibitem[\protect\citeauthoryear{Jetley, Gioachin, Mendes, Kale  \&
  Quinn}{Jetley et~al.}{2008}]{jetley}
Jetley P.,  Gioachin F.,  Mendes C.,  Kale L.~V.,   Quinn T.,  2008, in 2008
  IEEE International Symposium on Parallel and Distributed Processing. pp
  1--12, \mn@doi{10.1109/IPDPS.2008.4536319}

\bibitem[\protect\citeauthoryear{{Jogee}, {Scoville}  \& {Kenney}}{{Jogee}
  et~al.}{2005}]{jogee}
{Jogee} S.,  {Scoville} N.,   {Kenney} J.~D.~P.,  2005, \mn@doi [\apj]
  {10.1086/432106}, \href {http://adsabs.harvard.edu/abs/2005ApJ...630..837J}
  {630, 837}

\bibitem[\protect\citeauthoryear{{Kormendy}}{{Kormendy}}{2013}]{kormendy}
{Kormendy} J.,  2013, {Secular Evolution in Disk Galaxies}.
p.~1

\bibitem[\protect\citeauthoryear{{Kraljic}, {Bournaud}  \& {Martig}}{{Kraljic}
  et~al.}{2012}]{kraljic}
{Kraljic} K.,  {Bournaud} F.,   {Martig} M.,  2012, \mn@doi [\apj]
  {10.1088/0004-637X/757/1/60}, \href
  {http://adsabs.harvard.edu/abs/2012ApJ...757...60K} {757, 60}

\bibitem[\protect\citeauthoryear{{Kroupa}}{{Kroupa}}{2001}]{kroupa}
{Kroupa} P.,  2001, \mn@doi [\mnras] {10.1046/j.1365-8711.2001.04022.x}, \href
  {http://adsabs.harvard.edu/abs/2001MNRAS.322..231K} {322, 231}

\bibitem[\protect\citeauthoryear{{Kroupa}, {Tout}  \& {Gilmore}}{{Kroupa}
  et~al.}{1993}]{kroupa93}
{Kroupa} P.,  {Tout} C.~A.,   {Gilmore} G.,  1993, \mn@doi [\mnras]
  {10.1093/mnras/262.3.545}, \href
  {http://adsabs.harvard.edu/abs/1993MNRAS.262..545K} {262, 545}

\bibitem[\protect\citeauthoryear{{Laurikainen}, {Salo}  \&
  {Buta}}{{Laurikainen} et~al.}{2004}]{laurikainen}
{Laurikainen} E.,  {Salo} H.,   {Buta} R.,  2004, \mn@doi [\apj]
  {10.1086/383462}, \href {http://adsabs.harvard.edu/abs/2004ApJ...607..103L}
  {607, 103}

\bibitem[\protect\citeauthoryear{{Lee}, {Woo}, {Lee}, {Hwang}, {Lee}, {Sohn}
  \& {Lee}}{{Lee} et~al.}{2012}]{lee}
{Lee} G.-H.,  {Woo} J.-H.,  {Lee} M.~G.,  {Hwang} H.~S.,  {Lee} J.~C.,  {Sohn}
  J.,   {Lee} J.~H.,  2012, \mn@doi [\apj] {10.1088/0004-637X/750/2/141}, \href
  {http://adsabs.harvard.edu/abs/2012ApJ...750..141L} {750, 141}

\bibitem[\protect\citeauthoryear{{Li}, {Gadotti}, {Mao}  \& {Kauffmann}}{{Li}
  et~al.}{2009}]{li}
{Li} C.,  {Gadotti} D.~A.,  {Mao} S.,   {Kauffmann} G.,  2009, \mn@doi [\mnras]
  {10.1111/j.1365-2966.2009.15028.x}, \href
  {http://adsabs.harvard.edu/abs/2009MNRAS.397..726L} {397, 726}

\bibitem[\protect\citeauthoryear{{Lin}, {Cervantes Sodi}, {Li}, {Wang}  \&
  {Wang}}{{Lin} et~al.}{2014}]{lin}
{Lin} Y.,  {Cervantes Sodi} B.,  {Li} C.,  {Wang} L.,   {Wang} E.,  2014,
  \mn@doi [\apj] {10.1088/0004-637X/796/2/98}, \href
  {http://adsabs.harvard.edu/abs/2014ApJ...796...98L} {796, 98}

\bibitem[\protect\citeauthoryear{{{\L}okas}, {Ebrov{\'a}}, {del Pino},
  {Sybilska}, {Athanassoula}, {Semczuk}, {Gajda}  \& {Fouquet}}{{{\L}okas}
  et~al.}{2016}]{lokas}
{{\L}okas} E.~L.,  {Ebrov{\'a}} I.,  {del Pino} A.,  {Sybilska} A.,
  {Athanassoula} E.,  {Semczuk} M.,  {Gajda} G.,   {Fouquet} S.,  2016, \mn@doi
  [\apj] {10.3847/0004-637X/826/2/227}, \href
  {http://adsabs.harvard.edu/abs/2016ApJ...826..227L} {826, 227}

\bibitem[\protect\citeauthoryear{{L{\"u}tticke}, {Dettmar}  \&
  {Pohlen}}{{L{\"u}tticke} et~al.}{2000}]{lutticke}
{L{\"u}tticke} R.,  {Dettmar} R.-J.,   {Pohlen} M.,  2000, \aap, \href
  {http://adsabs.harvard.edu/abs/2000A\%26A...362..435L} {362, 435}

\bibitem[\protect\citeauthoryear{{Lynden-Bell} \& {Kalnajs}}{{Lynden-Bell} \&
  {Kalnajs}}{1972}]{lynden}
{Lynden-Bell} D.,  {Kalnajs} A.~J.,  1972, \mn@doi [\mnras]
  {10.1093/mnras/157.1.1}, \href
  {http://adsabs.harvard.edu/abs/1972MNRAS.157....1L} {157, 1}

\bibitem[\protect\citeauthoryear{{Martinet} \& {Friedli}}{{Martinet} \&
  {Friedli}}{1997}]{martinet}
{Martinet} L.,  {Friedli} D.,  1997, \aap, \href
  {http://adsabs.harvard.edu/abs/1997A\%26A...323..363M} {323, 363}

\bibitem[\protect\citeauthoryear{{Martinez-Valpuesta}, {Aguerri}  \&
  {Gonz{\'a}lez-Garc{\'{\i}}a}}{{Martinez-Valpuesta} et~al.}{2016}]{martinez}
{Martinez-Valpuesta} I.,  {Aguerri} J.,   {Gonz{\'a}lez-Garc{\'{\i}}a} C.,
  2016, \mn@doi [Galaxies] {10.3390/galaxies4020007}, \href
  {http://adsabs.harvard.edu/abs/2016Galax...4....7M} {4, 7}

\bibitem[\protect\citeauthoryear{{Mayer} \& {Wadsley}}{{Mayer} \&
  {Wadsley}}{2004}]{mayer}
{Mayer} L.,  {Wadsley} J.,  2004, \mn@doi [\mnras]
  {10.1111/j.1365-2966.2004.07202.x}, \href
  {http://adsabs.harvard.edu/abs/2004MNRAS.347..277M} {347, 277}

\bibitem[\protect\citeauthoryear{{Menon}, {Wesolowski}, {Zheng}, {Jetley},
  {Kale}, {Quinn}  \& {Governato}}{{Menon} et~al.}{2015}]{menon}
{Menon} H.,  {Wesolowski} L.,  {Zheng} G.,  {Jetley} P.,  {Kale} L.,  {Quinn}
  T.,   {Governato} F.,  2015, \mn@doi [Computational Astrophysics and
  Cosmology] {10.1186/s40668-015-0007-9}, \href
  {http://adsabs.harvard.edu/abs/2015ComAC...2....1M} {2, 1}

\bibitem[\protect\citeauthoryear{{Moetazedian}, {Polyachenko}, {Berczik}  \&
  {Just}}{{Moetazedian} et~al.}{2017}]{moetazedian}
{Moetazedian} R.,  {Polyachenko} E.~V.,  {Berczik} P.,   {Just} A.,  2017,
  \mn@doi [\aap] {10.1051/0004-6361/201630024}, \href
  {http://adsabs.harvard.edu/abs/2017A\%26A...604A..75M} {604, A75}

\bibitem[\protect\citeauthoryear{{Nair} \& {Abraham}}{{Nair} \&
  {Abraham}}{2010}]{nair}
{Nair} P.~B.,  {Abraham} R.~G.,  2010, \mn@doi [\apjl]
  {10.1088/2041-8205/714/2/L260}, \href
  {http://adsabs.harvard.edu/abs/2010ApJ...714L.260N} {714, L260}

\bibitem[\protect\citeauthoryear{{Okamoto}, {Isoe}  \& {Habe}}{{Okamoto}
  et~al.}{2015}]{okamoto}
{Okamoto} T.,  {Isoe} M.,   {Habe} A.,  2015, \mn@doi [\pasj]
  {10.1093/pasj/psv037}, \href
  {http://adsabs.harvard.edu/abs/2015PASJ...67...63O} {67, 63}

\bibitem[\protect\citeauthoryear{{Ostriker} \& {Peebles}}{{Ostriker} \&
  {Peebles}}{1973}]{ostriker}
{Ostriker} J.~P.,  {Peebles} P.~J.~E.,  1973, \mn@doi [\apj] {10.1086/152513},
  \href {http://adsabs.harvard.edu/abs/1973ApJ...186..467O} {186, 467}

\bibitem[\protect\citeauthoryear{{Pontzen}, {Tremmel}, {Roth}, {Peiris},
  {Saintonge}, {Volonteri}, {Quinn}  \& {Governato}}{{Pontzen}
  et~al.}{2017}]{pontzen}
{Pontzen} A.,  {Tremmel} M.,  {Roth} N.,  {Peiris} H.~V.,  {Saintonge} A.,
  {Volonteri} M.,  {Quinn} T.,   {Governato} F.,  2017, \mn@doi [\mnras]
  {10.1093/mnras/stw2627}, \href
  {http://adsabs.harvard.edu/abs/2017MNRAS.465..547P} {465, 547}

\bibitem[\protect\citeauthoryear{{Roberts}, {Huntley}  \& {van
  Albada}}{{Roberts} et~al.}{1979}]{roberts}
{Roberts} Jr. W.~W.,  {Huntley} J.~M.,   {van Albada} G.~D.,  1979, \mn@doi
  [\apj] {10.1086/157367}, \href
  {http://adsabs.harvard.edu/abs/1979ApJ...233...67R} {233, 67}

\bibitem[\protect\citeauthoryear{{Rodionov}, {Athanassoula}  \&
  {Peschken}}{{Rodionov} et~al.}{2017}]{rodionov}
{Rodionov} S.~A.,  {Athanassoula} E.,   {Peschken} N.,  2017, \mn@doi [\aap]
  {10.1051/0004-6361/201628319}, \href
  {http://adsabs.harvard.edu/abs/2017A\%26A...600A..25R} {600, A25}

\bibitem[\protect\citeauthoryear{{Romano-D{\'{\i}}az}, {Shlosman}, {Heller}  \&
  {Hoffman}}{{Romano-D{\'{\i}}az} et~al.}{2008}]{romano}
{Romano-D{\'{\i}}az} E.,  {Shlosman} I.,  {Heller} C.,   {Hoffman} Y.,  2008,
  \mn@doi [\apjl] {10.1086/593168}, \href
  {http://adsabs.harvard.edu/abs/2008ApJ...687L..13R} {687, L13}

\bibitem[\protect\citeauthoryear{{Sanders} \& {Huntley}}{{Sanders} \&
  {Huntley}}{1976}]{sanders}
{Sanders} R.~H.,  {Huntley} J.~M.,  1976, \mn@doi [\apj] {10.1086/154692},
  \href {http://adsabs.harvard.edu/abs/1976ApJ...209...53S} {209, 53}

\bibitem[\protect\citeauthoryear{{Scannapieco} \& {Athanassoula}}{{Scannapieco}
  \& {Athanassoula}}{2012}]{scannapieco}
{Scannapieco} C.,  {Athanassoula} E.,  2012, \mn@doi [\mnras]
  {10.1111/j.1745-3933.2012.01291.x}, \href
  {http://adsabs.harvard.edu/abs/2012MNRAS.425L..10S} {425, L10}

\bibitem[\protect\citeauthoryear{{Sellwood}}{{Sellwood}}{2014}]{sellwood}
{Sellwood} J.~A.,  2014, \mn@doi [Reviews of Modern Physics]
  {10.1103/RevModPhys.86.1}, \href
  {http://adsabs.harvard.edu/abs/2014RvMP...86....1S} {86, 1}

\bibitem[\protect\citeauthoryear{{S{\'e}rsic}}{{S{\'e}rsic}}{1963}]{sersic}
{S{\'e}rsic} J.~L.,  1963, Boletin de la Asociacion Argentina de Astronomia La
  Plata Argentina, \href {http://adsabs.harvard.edu/abs/1963BAAA....6...41S}
  {6, 41}

\bibitem[\protect\citeauthoryear{{Skibba} et~al.,}{{Skibba}
  et~al.}{2012}]{skibba}
{Skibba} R.~A.,  et~al., 2012, \mn@doi [\mnras]
  {10.1111/j.1365-2966.2012.20972.x}, \href
  {http://adsabs.harvard.edu/abs/2012MNRAS.423.1485S} {423, 1485}

\bibitem[\protect\citeauthoryear{{Soko\l{}owska}, {Capelo}, {Fall}, {Mayer},
  {Shen}  \& {Bonoli}}{{Soko\l{}owska} et~al.}{2017}]{sokowlowska}
{Soko\l{}owska} A.,  {Capelo} P.~R.,  {Fall} S.~M.,  {Mayer} L.,  {Shen} S.,
  {Bonoli} S.,  2017, \mn@doi [\apj] {10.3847/1538-4357/835/2/289}, \href
  {http://adsabs.harvard.edu/abs/2017ApJ...835..289S} {835, 289}

\bibitem[\protect\citeauthoryear{{Spergel} et~al.,}{{Spergel}
  et~al.}{2007}]{spergel}
{Spergel} D.~N.,  et~al., 2007, \mn@doi [\apjs] {10.1086/513700}, \href
  {http://adsabs.harvard.edu/abs/2007ApJS..170..377S} {170, 377}

\bibitem[\protect\citeauthoryear{{Spinoso}, {Bonoli}, {Dotti}, {Mayer}, {Madau}
   \& {Bellovary}}{{Spinoso} et~al.}{2017}]{spinoso}
{Spinoso} D.,  {Bonoli} S.,  {Dotti} M.,  {Mayer} L.,  {Madau} P.,
  {Bellovary} J.,  2017, \mn@doi [\mnras] {10.1093/mnras/stw2934}, \href
  {http://adsabs.harvard.edu/abs/2017MNRAS.465.3729S} {465, 3729}

\bibitem[\protect\citeauthoryear{{Springel} et~al.,}{{Springel}
  et~al.}{2008}]{springelAqua}
{Springel} V.,  et~al., 2008, \mn@doi [\mnras]
  {10.1111/j.1365-2966.2008.14066.x}, \href
  {http://adsabs.harvard.edu/abs/2008MNRAS.391.1685S} {391, 1685}

\bibitem[\protect\citeauthoryear{{Stinson}, {Seth}, {Katz}, {Wadsley},
  {Governato}  \& {Quinn}}{{Stinson} et~al.}{2006}]{stinson}
{Stinson} G.,  {Seth} A.,  {Katz} N.,  {Wadsley} J.,  {Governato} F.,   {Quinn}
  T.,  2006, \mn@doi [\mnras] {10.1111/j.1365-2966.2006.11097.x}, \href
  {http://adsabs.harvard.edu/abs/2006MNRAS.373.1074S} {373, 1074}

\bibitem[\protect\citeauthoryear{{Toomre}}{{Toomre}}{1964}]{toomre}
{Toomre} A.,  1964, \mn@doi [\apj] {10.1086/147861}, \href
  {http://adsabs.harvard.edu/abs/1964ApJ...139.1217T} {139, 1217}

\bibitem[\protect\citeauthoryear{{Toomre}}{{Toomre}}{1981}]{toomreB}
{Toomre} A.,  1981, in {Fall} S.~M.,  {Lynden-Bell} D.,  eds, Structure and
  Evolution of Normal Galaxies. pp 111--136

\bibitem[\protect\citeauthoryear{{Valenzuela} \& {Klypin}}{{Valenzuela} \&
  {Klypin}}{2003}]{valenzuela}
{Valenzuela} O.,  {Klypin} A.,  2003, \mn@doi [\mnras]
  {10.1046/j.1365-8711.2003.06947.x}, \href
  {http://adsabs.harvard.edu/abs/2003MNRAS.345..406V} {345, 406}

\bibitem[\protect\citeauthoryear{{Wadsley}, {Stadel}  \& {Quinn}}{{Wadsley}
  et~al.}{2004}]{wadsley}
{Wadsley} J.~W.,  {Stadel} J.,   {Quinn} T.,  2004, \mn@doi [\na]
  {10.1016/j.newast.2003.08.004}, \href
  {http://adsabs.harvard.edu/abs/2004NewA....9..137W} {9, 137}

\makeatother
\end{thebibliography}


\bsp	
\label{lastpage}
\end{document}